\documentclass[a4paper,fleqn,authoryear]{cas-sc}

\usepackage[authoryear]{natbib}

\usepackage{graphicx}
\usepackage{float}
\usepackage{algorithm}
\usepackage{algpseudocode}
\usepackage{enumitem}
\usepackage{array}
\usepackage{placeins}
\usepackage{capt-of}
\usepackage{hyperref}

\def\tsc#1{\csdef{#1}{\textsc{\lowercase{#1}}\xspace}}
\tsc{WGM}
\tsc{QE}

\AtBeginDocument{
\setlength{\abovedisplayskip}{3pt plus 1pt minus 1pt}
\setlength{\belowdisplayskip}{3pt plus 1pt minus 1pt}
\setlength{\abovedisplayshortskip}{3pt plus 1pt minus 1pt}
\setlength{\belowdisplayshortskip}{3pt plus 1pt minus 1pt}
\setlength{\jot}{1pt} 
}

\begin{document}
\let\WriteBookmarks\relax
\def\floatpagepagefraction{0.2}
\def\textpagefraction{.001}

\shorttitle{Li et al. (2026) \textit{OhmicFlow}: Forecasting transit passenger flow under extreme weather disruptions via Ohm's law}    

\shortauthors{Li et~al.}  

\title [mode = title]{\textit{OhmicFlow}: Forecasting transit passenger flow under extreme weather disruptions via Ohm's law}

%

\author[1]{Tianhao Li}[ orcid=0000-0003-3093-0447, ]
\ead{tianhao123@connect.hku.hk}
\credit{Writing - original draft, Methodology, Formal analysis, Data curation, Conceptualization, Visualization}

\author[1]{Xintian Liu}[ orcid=0009-0007-5871-7951, ]
\ead{xintianliu@connect.hku.hk}
\credit{Writing - original draft, Formal analysis}

\author[1,2,3]{Zhan Zhao}[ orcid=0000-0001-5170-9608, ]
\cormark[1]
\ead{zhanzhao@hku.hk}
\credit{Writing - review \& editing, Supervision, Methodology, Conceptualization}

\affiliation[1]{organization={Department of Urban Planning and Design, The University of Hong Kong},
                city={Hong Kong SAR},
                country={China}}

\affiliation[2]{organization={Urban Systems Institute, The University of Hong Kong},
                city={Hong Kong SAR},
                country={China}}

\affiliation[3]{organization={Musketeers Foundation Institute of Data Science, The University of Hong Kong},
                city={Hong Kong SAR},
                country={China}}

\cortext[1]{Corresponding author}

\begin{abstract}
As climate change intensifies, extreme weather events (EWEs) increasingly disrupt the supply-demand balance of mass transit systems. Accurate and reliable prediction of origin-destination (OD) passenger flow is essential for timely emergency response, but remains difficult under abnormal conditions. Purely data-driven models tend to overfit regular patterns and generalize poorly to rare events. Although existing physics-informed methods can enhance model robustness, they remain insufficient for transit systems under extreme weather disruptions, where passenger flow fluctuations are jointly shaped by the redistribution of latent demand as a direct result of EWEs (i.e., direct effects) and the increase in travel impedance due to EWE-induced supply contraction and congestion effects (i.e., indirect effects). To address these limitations, we first conceptualize the transit network as an electrical circuit, treating latent demand as voltage, travel impedance as resistance and passenger flow as current, and then propose the \textit{OhmicFlow} framework to jointly predict these variables under disruptions through Ohm's law. Specifically, a future-aware spatiotemporal backbone is used as an ammeter to predict disrupted passenger flow, and its replica is reused as a bypass voltmeter with impedance controlled in the inputs to infer latent demand counterfactually and in parallel. Travel impedance under foreseeable EWE disruptions is further modeled using a thermistor analogy, enabling dynamic estimation by coupling supply contraction with congestion effects. A multi-objective loss is incorporated to fit observed data while enforcing the Ohmic constraint. Empirical experiments based on 10 years of Shenzhen Metro data covering 17 EWEs show that \textit{OhmicFlow} consistently outperforms various baseline methods, achieving lower prediction errors across various chronological training settings while improving uncertainty calibration, robustness, transferability, and interpretability. Although this study focuses mainly on extreme weather disruptions, the proposed framework is general and can be adapted to other disruption scenarios in the future. 
\end{abstract}



\maketitle
\section{Introduction}
Mass transit systems are essential to efficient urban mobility and sustained economic activity across cities worldwide, underscoring the importance of maintaining their operational resilience. As climate change intensifies, the growing frequency of extreme weather events (EWEs), such as typhoons and rainstorms, has rendered these systems increasingly susceptible to severe disruptions \citep{Li2026Disentangling,Zhou2021Analyzing}. Such events can simultaneously impair service supply and alter travel demand, thereby generating substantial uncertainty and network-wide supply-demand imbalances. To strengthen transit system resilience under EWE-induced disruptions, accurate prediction of origin-destination (OD) passenger flows is crucial, as reliable forecasts can inform timely operational interventions and emergency response. However, this remains a highly challenging problem that has been largely underexplored.

Existing methods for passenger flow forecasting are mostly data-driven and often struggle to capture disruption-specific mechanisms. Although advanced deep learning architectures can model complex nonlinear patterns \citep{Zhang2023Deep,Jin2025SASTGCN}, they are largely designed for normal conditions and usually learn spatiotemporal correlations from passenger flow data alone \citep{Zhang2023Deep}. Even when external covariates (e.g., weather and operational metrics) are included, most methods rely on simple feature concatenation \citep{Hu2024Graph,Ding2026Origin}. As a result, these data-driven models inadequately represent key multivariate interactions in disrupted transit systems, especially the coupled constraints of supply, demand, and environments. In addition, high model complexity can worsen generalizability by amplifying sensitivity to distributional shifts and risking overfitting when trained on scarce and heterogeneous data from extreme events \citep{Chen2019Subway,Zhang2025Multi}. 

Although Physics-Informed Neural Networks (PINNs) have been introduced to enhance model robustness under disruptions \citep{Chew2025Physics}, existing PINN-based approaches to mobility modeling remain poorly suited to transit systems under extreme weather disruptions, where fluctuations in passenger flow are jointly determined by both direct and indirect effects of EWEs. Direct effects arise mainly from the redistribution of latent demand as people's travel plans change in response to EWEs (e.g., shifted commuting schedules to avoid a typhoon), which directly reshape transit passenger flows. Indirect effects typically manifest through an increase in travel impedance due to supply contraction (e.g., reduced train speed and service frequency for safety reasons) and resulting congestion effects (e.g., overcrowding and denied boarding in stations), which can in turn suppress passenger flow in affected OD pairs. This coupled relationship between passenger flow, latent demand, and travel impedance is rarely made explicit in current PINN formulations. For example, field-theoretic and gravity models \citep{Rong2023Origin,Simini2021deep,Wu2025Physics} can characterize aggregate mobility patterns across geographic regions, but generally overlook network capacity constraints and potential congestion effects within transit systems. Similarly, methods based on Macroscopic Fundamental Diagrams (MFDs) and fluid dynamics \citep{Shi2021physics,Li2025Embedding} can capture the interaction between vehicular flow and congestion-induced impedance through speed-density relationships, yet fail to represent latent travel demand or its mismatch with constrained system supply.

Unlike other types of disruptions (e.g., accidents), which are largely unpredictable, weather-induced disruptions are often partially foreseeable through weather forecasts and advance warnings, although the precise timing and magnitude of their impacts remain highly uncertain. In addition, EWEs typically persist for several hours or even days, with weather conditions evolving dynamically over the course of the event. Accordingly, a robust prediction model should incorporate available time-varying weather information while accounting for the inherent uncertainty in its effects on passenger flow. Furthermore, given the rarity and heterogeneity of EWEs, it is essential for the model to learn generalizable patterns from previously observed events and apply them to the prediction of future unseen events, an issue that has been largely overlooked in prior studies.

To bridge these gaps, we propose a simple yet physically intuitive modeling approach that conceptualizes the transit network as an electrical circuit. Just as electrons flowing from high to low potential driven by voltage and constrained by resistance, passengers move from origins to destinations driven by latent demand and constrained by travel impedance. Grounded in this isomorphism, we introduce Ohm's law as an explicit physical constraint, and develop \textit{OhmicFlow} as a unified framework that couples latent demand and travel impedance with passenger flow. Specifically, it employs a parallel dual-forward Ammeter-Voltmeter architecture, where the Ammeter predicts disrupted flow via factual signals of the main branch, while the Voltmeter counterfactually infers latent demand through a impedance-controlled bypass. Concurrently, it introduces a Positive Temperature Coefficient (PTC)-inspired module to model the dynamic impedance of the main branch by combining supply contraction and congestion effects. To provide a strong sensing capacity for the Ammeter and Voltmeter, we design a spatiotemporal deep learning module named TFT-GAT. Temporally, a future-aware Temporal Fusion Transformer (TFT) \citep{Lim2021Temporal} integrates forward-looking covariates (e.g., typhoon warnings and planned service restrictions) into multi-scale temporal dependencies to encode expected disruption attributes. Spatially, a Graph Attention Network (GAT) \citep{Velickovic2018Graph} performs high-order message passing over the hidden state information of TFT, mitigating over-smoothing \citep{Zhang2025Efficient} and computational bottlenecks \citep{Shao2022Pre} inherent to step-wise graph attention. Although designed for extreme weather disruptions, this framework can be extended to other similar disruption scenarios, such as those induced by major public events (e.g., concerts or sports competitions) \citep{Liang2024Exploring}, when future event attributes and operation signals are available. 

In summary, the core contributions of this paper are threefold:

\begin{itemize}[noitemsep]
\item \textbf{Physics-Informed Formulation:} We introduce a new modeling perspective that conceptualizes the transit network as an electrical circuit, treating latent demand as voltage, travel impedance as resistance, and passenger flow as current. By establishing the circuit isomorphism with Ohm's law as an explicit constraint, we bridge the theoretical gap in the joint modeling of passenger flow and its physical drivers in public transit systems.
\item \textbf{Model Design:} We propose \textit{OhmicFlow}, a modeling framework that integrates a data-driven Ammeter with physical structure inspired by the Voltmeter and PTC. Leveraging unified embeddings and joint loss function, the Ammeter predicts disrupted flow, the Voltmeter counterfactually infers latent demand, while PTC concurrently captures dynamic travel impedance. The proposed model combines spatiotemporal learning capacity with the robust physical constraints of Ohm's law. Although we use TFT-GAT as the default sensing backbone for both meters, \textit{OhmicFlow} can accommodate any deep learning model.
\item \textbf{Empirical Validation:} Leveraging a 10-year, 17-event real-world dataset from Shenzhen Metro, we evaluate \textit{OhmicFlow} under different data splitting schemes. The results show consistent superiority over existing baselines in both predictive accuracy and uncertainty mitigation, and the performance remains robust across heterogeneous OD pairs and varied perturbation settings. Ablation studies further reveal that the efficacy of the Ohmic constraint scales with the disruption severity. Moreover, \textit{OhmicFlow} also exhibits plug-and-play extensibility to enhance existing backbones, while providing interpretable insights for emergency management.
\end{itemize}

The outline of this paper is as follows. Section~\ref{sec:literature} reviews data-driven and physics-informed methods for passenger flow forecasting under disruption. Section~\ref{sec:preliminaries} formalizes the transit-circuit isomorphism and the problem statement. Section~\ref{sec:method} details the \textit{OhmicFlow} framework and its three modules (i.e., Ammeter, Voltmeter, PTC). Section~\ref{sec:case} describes the case study, including data, settings, and baselines. Sections~\ref{sec:results} and \ref{sec:conclusion} present empirical results and conclusions.

\section{Literature review}\label{sec:literature}

\subsection{Data-driven passenger flow forecasting under disruption}

Methods for forecasting transit passenger flow have evolved from classical statistical models \citep{Cardozo2012Application,Ding2018Using}, machine learning approaches \citep{Leng2013Probability,Sun2015novel}, and feed-forward neural networks \citep{Wei2012Forecasting} to spatiotemporal graph neural networks \citep{Li2018Diffusion,Wu2019Graph,Yu2018Spatio} and attention-based architectures \citep{Lim2021Temporal,Xu2021Spatial,Zheng2020GMAN}. The latest models can achieve high prediction accuracy under normal conditions, in part due to the high periodicity of human mobility patterns. However, disruptive events such as EWEs can drastically reshape both passenger demand and operational supply of public transit systems in a short period, making it a formidable challenge to predict passenger flows under such abnormal conditions. Recent work on passenger flow forecasting under disruptions can be categorized into two paradigms: (i) modeling disruptions implicitly from flow dynamics alone, and (ii) explicitly integrating event features or proxies as exogenous inputs.

When disruption signals are absent or unreliable, a stream of studies leveraged flow anomalies as implicit disruption cues to enhance model sensitivity. From a statistical perspective, \cite{Chen2020Subway} decoupled mean and volatility modeling with an updated Autoregressive Integrated Moving Average (ARIMA) framework, where large residuals inflated conditional variance and widened final prediction intervals. Similarly, \cite{Noursalehi2018Real} introduced Kalman filtering into a dynamic factor model, using extracted prediction residuals to correct hidden states and thus enable model adaptation to unknown disruption scenarios. In deep learning, \cite{Huang2023Extreme} overcame the peak smoothing problem with an extreme value scaling mechanism that decomposed flow into a normal baseline and an extreme deviation multiplier learned by a dedicated Gated Recurrent Units (GRU). \cite{Ren2022TBSM} combined K-Nearest Neighbors (KNN) retrieval with a deep agent to adaptively reweight absolute status and transient fluctuation during flow bursty periods, improving the matching of historical surge patterns. However, reliance on a single source of flow observations inherently constrains theoretical performance. Because these models do not explicitly account for disruption context, they largely extrapolate from historical patterns. As a result, they tend to overfit the flow dynamics observed in the training data without adequately distinguishing among heterogeneous shocks, thereby limiting their capacity to learn more generalizable representations \citep{Liu2020Impacts}.

To address these issues, another stream of work explicitly inject multi-source event features or proxies as external context. The most direct approach fuses numerical event features with spatiotemporal flow data. For instance, \cite{Noursalehi2018Real} encoded the spatiotemporal scope of planned events into dummy variable matrices to forcefully intervene in model predictions, while \cite{Tang2019Forecasting} and \cite{Roy2021Predicting} directly fed external variables, such as rainfall intensity and hurricane proximity, into Support Vector Regression (SVR) and Long Short-Term Memory (LSTM) networks. For unobserved events, \cite{Li2017Forecasting} extracted lagged inflows from other stations as proxies for disruptions, utilizing a Multi-Scale Radial Basis Function (MSRBF) network to capture intense localized outflow surges. To prevent critical event effects from being diluted, some studies separate regular flow patterns from event-induced deviations. For example, \cite{Yu2017Deep} employed Stacked Autoencoders (SAE) to extract a denoised perturbation vector from traffic accident attributes, merging it downstream with an LSTM-derived regular dynamic baseline. \cite{Xue2022Forecasting} first quantified perturbations from network-wide inflow and social media intent via Convolutional Neural Networks (CNNs) and SAE, and then subtracted them from historical observations to learn pure inherent trends. Recently, the integration of event features has been extended to deep semantic extraction of unstructured data. \cite{Zou2024Real} mapped textual incident reports into tangible 3D blockage tensors and injected them into graph convolutions with attention, while \cite{Liang2024Exploring} extracted structured event contexts from fragmented activity notices and coupled them with flow sequences as prompts for LLM-based reasoning. Despite the advances, these data-driven approaches tend to struggle with out-of-distribution (OOD) samples that deviate significantly from historical training distributions \citep{Yang2024Generalized}, frequently yielding physically implausible predictions. Moreover, most studies focus on real-time event inputs and overlook forward-looking contextual information (e.g., planned suspensions or public alerts) that can induce behavioral shifts even before the event fully unfolds, preventing the model from reconstructing the entire disruption lifecycle.
 
\subsection{PINNs for mobility flow modeling}

To overcome the limitations of data-driven methods, recent studies have explored PINNs to embed validated physical priors for flow estimation, which reduces reliance on dense high-quality data and improves model robustness under disruption. Depending on the physical principles adopted, existing applications of PINNs for flow modeling can be broadly categorized into three types: fluid dynamics, spatial interaction theory, and field theory.

First, fluid dynamics-inspired models typically conceptualize traffic as compressible fluids and enforce mechanistic constraints by integrating MFDs with continuity Partial Differential Equations (PDEs), such as the Lighthill-Whitham-Richards (LWR) model. For example, \cite{Shi2022physics} replaced the rigid MFD formula with a neural surrogate to map density to flow and then penalized predictions that violated the LWR equation, while \cite{Zhang2024Physicsinformed} calibrated MFD parameters from sparse data and embedded them into PDE residuals to control overfitting. To lower the burden of solving PDEs, \cite{Li2025Embedding} discretized the fluid continuity equation into learnable spatiotemporal feature tensors.
Second, the spatial interaction theory, rooted in radiation and universal gravitation, has also inspired mobility flow modeling. \cite{Simini2012universal} derived a radiation model for macroscopic mobility, and \cite{Simini2021deep} proved that the gravity model is equivalent to a softmax linear layer before generalizing it into a deep gravity network. \cite{Rong2023Origin} generalized this concept to Generative Adversarial Networks (GANs) and embedded gravity equations into OD synthesis. More recently, \cite{liangGeneratingSparseOrigin2024} extended this spatial-interaction perspective to shared mobility networks by using probabilistic GNNs to learn how OD contexts and spatial proximity shape mobility flows. For short-term passenger flow forecasting, \cite{Shen2021Hybrid} fused gravity-based estimates with CNN predictions.
Third, field-theoretic approaches usually employed fields to describe the underlying driving forces of macroscopic spatial mobility. \cite{Mazzoli2019Field} first mapped aggregated OD flows onto a continuous vector field satisfying the divergence theorem, \cite{Wang2023Traffic} then decomposed traffic flow graphs into potential energy polytrees for predictable evolution. For flooding-induced disruption, \cite{Lin2026Field} modeled cascading failures through field dynamics to regularize mobility prediction. 

Despite their potential, these physical priors are not well-suited to passenger flow forecasting in disrupted transit systems. Fluid dynamics and MFD-based models are mainly designed for vehicular flow already moving within the road network, while field theory and spatial interaction models are better suited to aggregate regional mobility than supply-constrained transit systems. More importantly, they do not explicitly capture the key disruption mechanism in public transit systems under EWEs, where passenger flow is jointly determined by the redistribution of latent demand (direct effects) and increase in travel impedance (indirect effects), the latter of which can also interact with passenger flow through congestion effects. Some physical models appear to contain similar multivariate relations, but they only capture parts of this mechanism. For example, gravity models represent demand and impedance through production-attraction and distance decay, but they do not capture how flow is affected by degraded transit service and how it feeds back into travel impedance. Similarly, MFDs describe the relations among flow, density, and speed of road vehicles, but do not directly model latent demand as the driving force behind passenger flow. 

To bridge the above research gaps, we propose to conceptualize the transit network as an electrical circuit. Under this analogy, the latent demand, travel impedance and passenger flow are linked through Ohm's law, allowing the prediction task to be reformulated from flow regression to multivariate inference. Specifically, each OD pair can be represented as a PTC thermistor, whose impedance depends on both the transit service supply and station-level congestion effects. Building on this, we develop the \textit{OhmicFlow} framework, which integrates a future-aware spatiotemporal backbone with a circuit-inspired physical architecture, using shared embeddings and a multi-objective loss to align data-driven and physical modules. This design promotes knowledge consistency and mutual validation between modules, thereby improving model robustness.

\section{Preliminaries}\label{sec:preliminaries}
\subsection{Transit-circuit isomorphism and assumptions}

\begin{figure}
    \centering
    \includegraphics[width=0.95\linewidth,height=0.4\textheight,keepaspectratio]{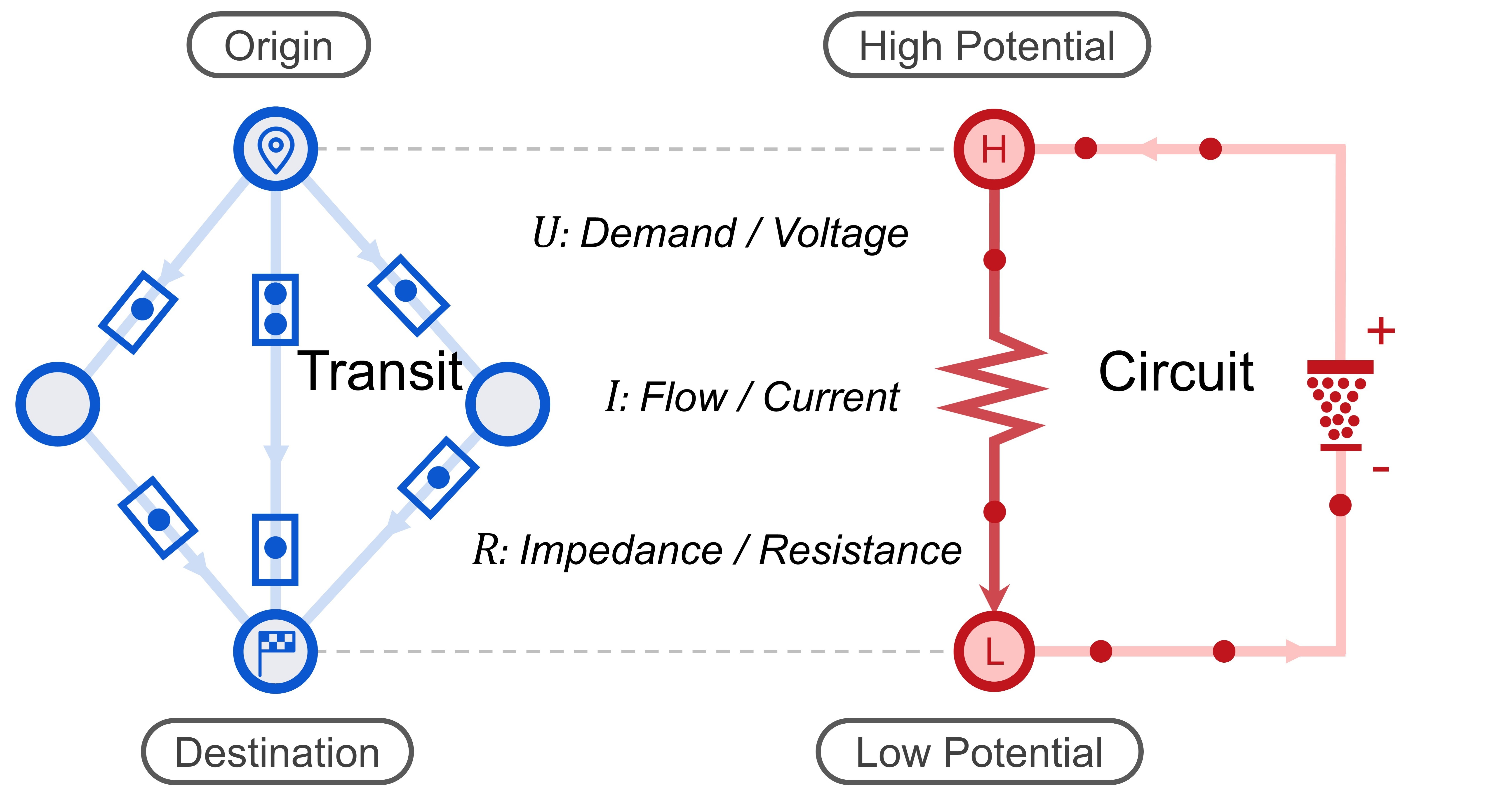}
    \vspace{-5pt}
    \captionof{figure}{Analogy between transit and circuit networks.}
    \label{fig:isomorphism}
\end{figure}

As shown in Fig.~\ref{fig:isomorphism}, transit networks and electrical circuits exhibit structural isomorphism. Transit stations and OD pairs correspond to circuit nodes and branches, and passenger movement parallels electron flow. The dynamic analogy is also consistent: both passenger flow and electric current ($I$) are driven by a source term ($U$, latent demand vs. voltage), constrained by a resistance term ($R$, travel impedance vs. electrical resistance), and subject to nodal conservation (the dispersion of station inflow into OD flow vs. Kirchhoff's Current Law). Kirchhoff's Current Law (KCL), a core principle of circuit analysis, states that at any node the total incoming current equals the total outgoing current. Transit systems exhibit an analogous nodal property: inflow entering a station is redistributed through OD pairs toward different destinations, providing a natural basis for transit-circuit isomorphism.

To operationalize this isomorphism, we introduce three assumptions:

\begin{enumerate}[noitemsep]
\item Inspired by Thevenin's theorem \citep{Johnson2003}, each transit OD pair can be represented as an equivalent branch consisting of a voltage source in series with a resistor. In this analogy, the voltage source corresponds to latent OD demand ($U$), while the resistor corresponds to OD travel impedance ($R$).
\item Within each time window $\Delta t$, the isomorphic system is quasi-steady, and the OD passenger flow $I$ acting as the branch current should satisfy Ohm's law:$I = U / R$.
\item Under normal operational conditions (i.e., no indirect effects), the travel impedance is normalized to $R=1$, and the latent demand is translated losslessly into OD flow. This is the state of minimum impedance; therefore, $R\geq 1$ always holds.
\end{enumerate}

\subsection{Notations and definitions} 

Table~\ref{tab:notations} summarizes the key notations used throughout the paper. We formulate the problem of passenger flow forecasting under disruption in a multi-step encoder-decoder setting: for each OD pair $k=(i,j)$, the model uses a historical window $\mathcal{T}_h=[t-\Delta T+1,\,t]$ to predict passenger flow over a future window $\mathcal{T}_f=[t+1,\,t+\Delta T]$, where $\Delta T$ is the common window length. For readability, key variables are grouped into five categories.

\textbf{(1) Disruption-related variables.}
We consider two main types of disruption-related variables: weather state $W_t$ and travel time $\tau_{k,t}$. $W_t$ characterizes the EWE that affects both latent demand and travel impedance, whereas $\tau_{k,t}$ captures the EWE-induced increase in travel impedance as a result of both supply contraction and congestion effects \citep{Li2026Disentangling}. In the circuit analogy, $W_t$ acts as an ambient state that modulates resistance and voltage, while $\tau_{k,t}$ serves as a resistivity-like factor contributing dominantly to resistance formation. Although these variables are not perfectly known in advance, they can be approximated from valuable early warning information over $\mathcal{T}_f$, such as weather forecasts, operation announcements, historical responses to similar events, and anticipated service adjustments. We therefore treat them as forward-looking disruption signals, where $\{W_t\}_{t\in\mathcal{T}_f}$ represents the anticipated weather state and $\{\tau_{k,t}\}_{t\in\mathcal{T}_f}$ represents the anticipated disrupted travel time. The sensitivity of \textit{OhmicFlow} to perturbations in these signals is further examined in Sec.~\ref{sec:robustness}.

\textbf{(2) Flow-related variables.}
We define passenger flow $I_{k,t}$ as the branch current within the transit-circuit isomorphism, representing the number of trips for OD pair $k$ at time $t$. Specifically, $I_{k,t}$ is organized into two views: (i) historical observed flow $\{I_{k,t}\}_{t\in\mathcal{T}_h}$ for the past time window; and (ii) expected normal flow $\{I^b_{k,t}\}_{t\in\mathcal{T}_f}$ as a reference for the future time window, where $I^b_{k,t}$ is taken from the most recent week with no disruption based on the same day of week and time of day as $t$.

\textbf{(3) Latent physical covariables.}
To describe the physical co-variation of OD flow, we conceptualize latent demand $U_{k,t}$ as voltage and travel impedance $R_{k,t}$ as resistance. Both are unobservable and must be inferred by the model for the future horizon $\mathcal{T}_f$. In particular, $R_{k,t}$ is modulated from single-step $\tau_{k,t}$ integrating contextual factors (e.g., $W_t$) to convert raw disruptions into a computable impedance state.

\textbf{(4) OD-centered local graph.}
For each OD pair $k=(i,j)$, we construct an OD-centered local graph \citep{liangGeneratingSparseOrigin2024} to capture disrupted spatial proximity. Since the transit network may evolve over time, we use $G_p=(V_p,E_p,A_p)$ to denote the network topology in time period $p$, where $V_p$ is the station set, $E_p$ the edge set, and $A_p$ the travel time matrix (e.g., based on the shortest path). Given an OD pair $k$, its neighbor set, denoted as $\mathcal{N}_{p,k}$, can be defined as the 10 nearest OD pairs $k'=(i',j')$ with lowest $A_{p,i,i'}+A_{p,j,j'}$.

\textbf{(5) Contexts.}
Static context $f_{i,p},f_{k,p}$ describes station/OD infrastructure attributes within period $p$, such as the number of serving transit lines, network centrality, minimum transfer count, shortest-path travel time, and path redundancy. Dynamic context $H_t$ describes temporal states, such as time-of-day and day-of-week.

\begin{table}
\centering
\small
\caption{Notations and definitions with circuit analogy}
\label{tab:notations}
\begin{tabular}{lp{8.5cm}p{3.5cm}}
\hline
\textbf{Notation} & \textbf{Description} & \textbf{Circuit Analogy} \\
\hline
$p$ & Index of period with invariant transit network topology & -- \\
$t$ & Index of time interval (e.g., 1 hour) & -- \\
$V_p$ & Set of transit stations in period $p$ & -- \\
$E_p$ & Set of directed edges (station connections) in period $p$ & -- \\
$A_p \in \mathbb{R}^{|V_p| \times |V_p|}$ & Topological distance matrix (shortest hop-counts) in period $p$ & -- \\
$G_p = (V_p, E_p, A_p)$ & Transit network graph in period $p$ & Circuit \\
$i, j \in V_p$ & Indices of origin and destination station & Node \\
$k = (i, j)$ & Index of an OD pair & Branch \\
$\mathcal{N}_{p,k}$ & 10 nearest neighbor OD pairs for $k$ in period $p$ & Coupled branch set \\
$U_{k,t}$ & Latent demand for OD pair $k$ at time $t$ & Voltage \\
$I_{k,t}$ & Passenger flow for OD pair $k$ at time $t$ & \multirow{3}{*}{Current} \\
$I_{k,t}, t\in\mathcal{T}_h$ & Historical observed flow for OD pair $k$ at time $t$& \\
$I^b_{k,t}, t\in\mathcal{T}_f$ & Expected normal flow for OD pair $k$ at time $t$ & \\
$\tau_{k,t}$ & Actual travel time for OD pair $k$ at time $t$ & \multirow{3}{*}{Resistivity} \\
$\tau_{k,t}, t\in\mathcal{T}_f$ & Anticipated disrupted travel time for OD pair $k$ at time $t$ & \\
$\widetilde{\tau}_{k,t}, t\in\mathcal{T}_f$ & Normal travel time for OD pair $k$ at time $t$ & \\
$\boldsymbol{\epsilon}_{k,t}$ & OD efficiency representation combining $e_{k,t}$ and $\Delta e_{k,t}$ & \multirow{3}{*}{Conductivity} \\
$e_{k,t}$ & Travel efficiency for OD pair $k$ at time $t$ & \\
$\Delta e_{k,t}$ & Relative efficiency for OD pair $k$ at time $t$ & \\
$R_{k,t}$ & Travel impedance for OD pair $k$ at time $t$ & Resistance \\
$R_{k,t}^b$ & Base supply-related impedance for OD pair $k$ at time $t$ & \hspace{0.5em}Base resistance \\
$\mathbf s_{k,t}$ & OD Supply representation for OD pair $k$ at time $t$ & \hspace{1em}\multirow{2}{*}{NT conductivity} \\
                                                         $S_{k,t}$ & OD Supply index for OD pair $k$ at time $t$ & \\
$f_{i,p}, f_{k,p}$ & Static infrastructural features (station/OD) in period $p$ & \hspace{1em}Material property\\
$W_t$ & Weather conditions (e.g., wind power, rainfall) at time $t$ & \hspace{1em}\multirow{2}{*}{Ambient state} \\
$H_t$ & Temporal contexts (e.g., time of day, day of week) at time $t$ & \\
$C_{i,t}$ & Congestion-induced amplification of station $i$ at time $t$ & \hspace{0.5em}Joule heating \\
$\alpha_{i,t}$ & Congestion sensitivity coefficient of station $i$ at time $t$ & \hspace{1em}PTC coefficient \\
$\eta_{i,p}$ & Historical peak inflow of station $i$ observed in period $p$ & \hspace{1em}Material property \\
$I_{i,t}^{\text{in}}$ & Total inflow of station $i$ at time $t$ & \hspace{1em}Total current injection \\

\hline
\end{tabular}
\vspace{1ex} \\
\raggedright
\footnotesize \textit{Note:} NT represents normal temperature. Indented entries denote sub-items of the preceding non-indented entry.
\end{table}

\subsection{Problem statement}
For each OD pair $k=(i,j)$, we define historical temporal inputs
$\mathbf{x}_{k}^{(h)}=\{I_{k,t},W_{t},\tau_{k,t},H_{t}\}_{t\in\mathcal{T}_h}$,
forward-looking temporal inputs
$\mathbf{x}_{k}^{(f)}=\{W_{t},\tau_{k,t},H_{t},I^b_{k,t}\}_{t\in\mathcal{T}_f}$,
and static inputs
$\mathbf{x}_{k}^{(s)}=\{f_{i,p},f_{k,p},\mathcal{N}_{p,k}\}$.
Building on the above definitions, we formulate a physics-informed passenger flow forecasting task over $\mathcal{T}_f$. The objective is to learn a parametric mapping that jointly infers future flow and its latent physical drivers:
\begin{equation}
\left[\hat{\mathbf{I}}_k^{\mathcal{T}_f},\hat{\mathbf{U}}_k^{\mathcal{T}_f},\hat{\mathbf{R}}_k^{\mathcal{T}_f}\right]
=\mathcal{F}_{\theta}\left(\mathbf{x}_{k}^{(h)},\mathbf{x}_{k}^{(f)},\mathbf{x}_{k}^{(s)}\right),
\end{equation}
where $\theta$ are trainable parameters and bold symbols denote sequences over $\mathcal{T}_f$. Unlike conventional mobility flow prediction formulations that treat flow as the sole prediction target, our formulation casts the problem as joint inference of $\hat{I}$, $\hat{U}$, and $\hat{R}$ under Ohmic constraints.

\section{Methodology}\label{sec:method}

\subsection{Framework Overview}
\begin{figure}
    \centering
    \includegraphics[width=1.0\linewidth,height=0.8\textheight,keepaspectratio]{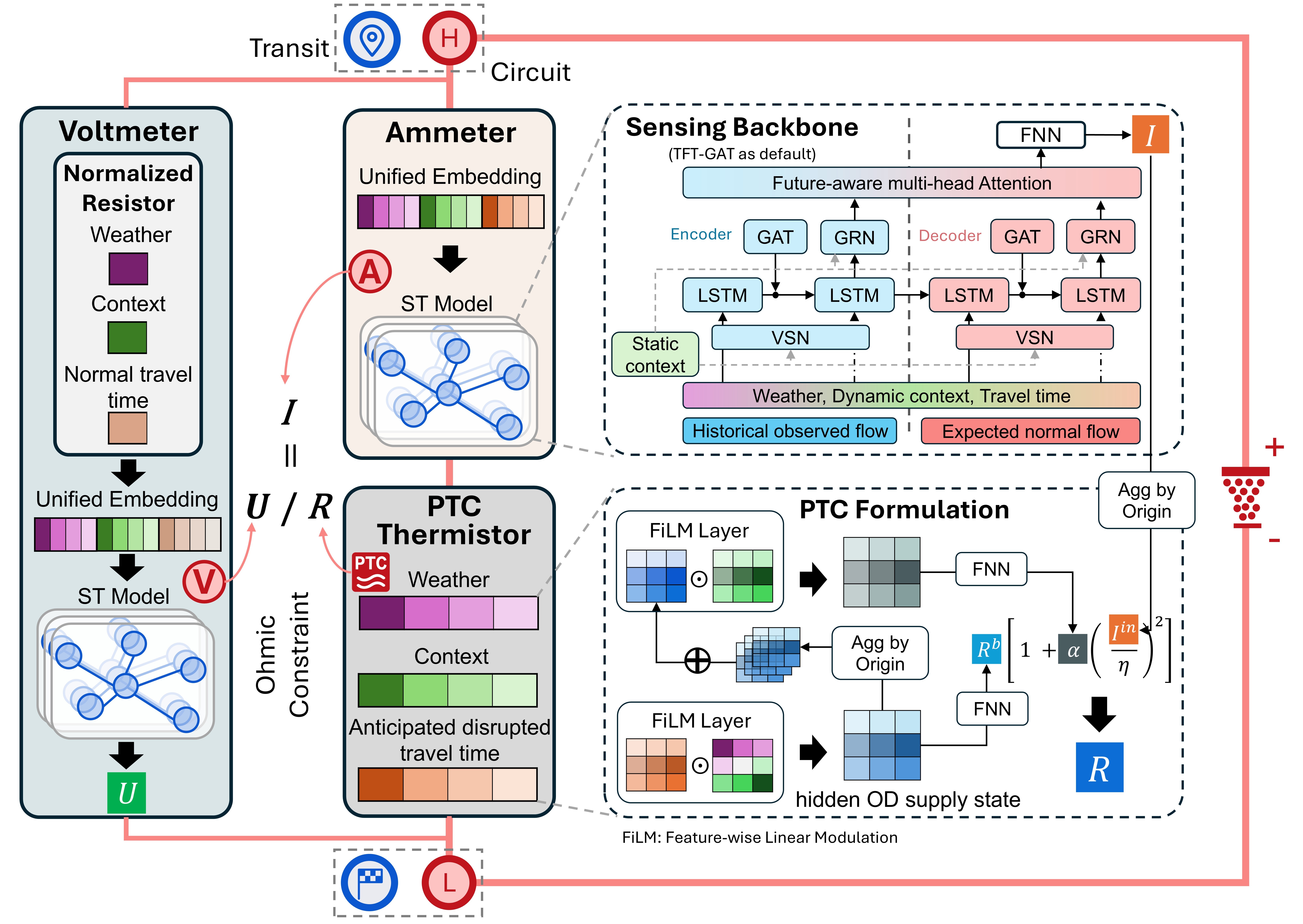}
    \vspace{-10pt}
    \captionof{figure}{Overall architecture of \textit{OhmicFlow} framework.}
    \label{fig:most}
\end{figure}

Fig.~\ref{fig:most} presents the proposed \textit{OhmicFlow} framework, which couples spatiotemporal learning with physical mechanisms. Structurally, it consists of a main branch in which an ammeter and a PTC thermistor are connected in series, together with a parallel voltmeter bypass. The ammeter uses a TFT-GAT sensing backbone to estimate disrupted passenger flow $\hat{I}$ from historical observations and future disruption signals. The voltmeter shares the backbone but replaces anticipated disrupted travel time with expected normal travel time (assuming no disruption) to counterfactually infer latent demand $\hat{U}$ under normal impedance conditions. The PTC thermistor estimates travel impedance $\hat{R}$ by first deriving base impedance from modulated hidden supply states and then scaling it by origin-level congestion effects based on modulated station sensitivity and aggregated inflow. Finally, the three modules are jointly trained and coupled through Ohm's law to produce physics-aware passenger flow estimates with improved robustness and physical consistency.

\subsection{Ammeter-equivalent spatiotemporal learning for flow forecasting} \label{sec:ammeter}

In the circuit analogy, the data-driven predictor of OD passenger flow plays the role of an ammeter, which is expected to generate accurate and robust forecasts under disruption. In this study, it includes two components: a unified embedding module and a future-aware sensing backbone. The former provides a shared semantic space for integrating information from other physical modules, ensuring representation consistency; and the latter captures spatiotemporal dependencies and anticipatory disruption cues to estimate disrupted passenger flow.

\subsubsection{Unified embedding for data-physics alignment}

As a core component of the ammeter, we introduce a unified embedding layer before the spatiotemporal sensing backbone. For each channel $c\in\{s,h,f\}$, numeric variables are embedded through linear projection while categorical variables are embedded by lookup tables. Both operations are uniformly denoted by $\phi(\cdot)$ for notation consistency. The channel-level embedding is then
\begin{equation}
\mathbf{e}^{(c)}=\mathrm{Concat}\!\left(\{\phi(x^{(c)}_{n})\}_{n=1}^{N_c}\right)\in\mathbb{R}^{d \times N_c},
\end{equation}
where $d$ is the state size, $x^{(c)}_{n}$ is the $n$-th variable in channel $c$, and $N_c$ is the total number of numeric and categorical variables in that channel. For temporal channels ($c\in\{h,f\}$), this mapping is applied at each time step.

Notably, the same embeddings are reused by the voltmeter and PTC to keep semantic representations synchronized across modules. This design strengthens the representations learned by the ammeter while providing a consistent semantic basis for coupled estimation across all modules.

\subsubsection{Sensing backbone powered by future-aware TFT-GAT}

Given aligned embeddings, the sensing backbone estimates disrupted passenger flow through feature enhancement, spatial semantic diffusion, and forward-looking temporal interaction.

\textbf{(1) Feature enhancement via GRN and VSN.} 

Under extreme weather disruptions, passenger flow signals show sharp nonlinear shifts and substantial exogenous noise. Following the TFT design \citep{Lim2021Temporal}, we combine the Gated Residual Network (GRN) and Variable Selection Network (VSN) to address these challenges. The GRN adaptively models nonlinear disruption patterns while preserving stable representations through residual connections, whereas the VSN performs instance-wise variable reweighting to suppress noisy or irrelevant inputs. Both modules use static context to inform dynamic representations, improving the robustness of sequence encoding to heterogeneous OD semantics.

Specifically, GRN is defined as
\begin{equation}
\mathrm{GRN}\!\left(\mathbf{c},\mathbf{s}\right)=\mathrm{LayerNorm}\!\left(\mathbf{c}+\mathrm{GLU}\!\left(\mathbf{W}_2\,\mathrm{ELU}\!\left(\mathbf{W}_1[\mathbf{c};\mathbf{s}]+\mathbf{b}_1\right)+\mathbf{b}_2\right)\right),
\end{equation}
where $\mathbf{c}$ denotes the representation of a dynamic sequence, $\mathbf{s}$ denotes the representation of static context, $\mathrm{ELU}(\cdot)$ denotes the Exponential Linear Unit \citep{Clevert2016Fast} that provides smooth and robust nonlinear activation, and $\mathrm{GLU}(\cdot)$ denotes the Gated Linear Unit \citep{Dauphin2017Language}, which first computes gate coefficients in $[0,1]$ and then applies element-wise modulation to control information pass-through. When no context is used, we set $\mathbf{s}=\mathbf{0}$.

For temporal channel $c\in\{h,f\}$, $\mathrm{VSN}\!\left(\mathbf{e}^{(c)},\mathbf{e}^{(s)}\right)$ then executes three steps:
\begin{equation}
\begin{aligned}
&\mathbf{g}_{n,t}^{(c)}=\mathrm{GRN}_{n}\!\left({\mathbf{e}}_{n,t}^{(c)}\right),\qquad n=1,\dots,N_c,\\
&\omega_{n,t}^{(c)}=\mathrm{softmax}_{n}\!\left(\mathrm{GRN}_{\omega}\!\left([{\mathbf{e}}_{1,t}^{(c)};\ldots;{\mathbf{e}}_{N_c,t}^{(c)}],\boldsymbol{\psi}\right)\right),\\
&{\widetilde{\mathbf{e}}}_{t}^{(c)}=\sum_{n=1}^{N_c}\omega_{n,t}^{(c)}\,\mathbf{g}_{n,t}^{(c)}.
\end{aligned}
\end{equation}

Here, $\boldsymbol{\psi}=\mathrm{Dense}\!\left(\mathbf{e}^{(s)}\right)$, where $\mathrm{Dense}(\cdot)$ projects static embeddings from $\mathbb{R}^{d \times N_s}$ to $\mathbb{R}^{d}$ for dimension-aligned context conditioning. The fused vector ${\widetilde{\mathbf{e}}}_{t}^{(c)}$ is used as the enhanced dynamic inputs for OD pair $k$ at time $t$.

\textbf{(2) Spatial modeling via interleaved GAT-LSTM layers.} 

Instead of stepwise spatial convolution, we place graph attention between stacked LSTM layers so that spatial information is injected at the semantic level rather than at every time step. At layer $l$, GAT aggregates the final hidden states from the previous LSTM layer across OD neighbors, and the resulting spatial summary is then used to initialize the current-layer recurrence:
\begin{equation}
\begin{aligned}
&\mathbf{z}_{k}^{\langle l\rangle} =
\begin{cases}
\boldsymbol{\psi}_{k}, & l=1,\\
\sum_{k'\in\mathcal{N}_{p,k}}\alpha_{k,k'}^{\langle l\rangle}\mathbf{W}_{G}^{\langle l\rangle}\mathbf{h}_{k'}^{\langle l-1\rangle}, & l>1,
\end{cases}\\
&\boldsymbol{\chi}_{k}^{\langle l\rangle},\mathbf{h}_{k}^{\langle l\rangle} = \mathrm{LSTM}\!\left(\boldsymbol{\chi}_{k}^{\langle l-1\rangle},\mathbf{z}_{k}^{\langle l\rangle}\right).
\end{aligned}
\end{equation}
where $\alpha_{k,k'}^{\langle l\rangle}$ is the attention coefficient, $\mathbf{h}_{k'}^{\langle l-1\rangle}$ is the final hidden state of neighbor $k'$, and $\boldsymbol{\psi}_{k}$ is the densified static context. For continuity, we set $\boldsymbol{\chi}_{k}^{\langle 0\rangle}=\widetilde{\mathbf{e}}^{(c)}$. Specifically, the final hidden state of the last encoder LSTM layer initializes the first decoder LSTM layer. This sequence-level spatial injection keeps message passing in a high-level semantic space, which improves efficiency and mitigates over-smoothing compared with per-step graph diffusion.

\textbf{(3) Temporal modeling with future-aware attention.} 

After the final layer $L$, we first apply a post-LSTM GRN refinement, then perform attention over the full horizon to capture long-range dependencies and anticipatory effects from forward-looking inputs:

\begin{equation}
\begin{aligned}
&\widetilde{\boldsymbol{\chi}}_{k}=\mathrm{GRN}_{\mathrm{post}}\!\left(\boldsymbol{\chi}_{k}^{\langle L\rangle},\boldsymbol{\psi}_{k}\right),\\
&\mathbf{Q},\mathbf{K},\mathbf{V}=\widetilde{\boldsymbol{\chi}}_{k}\mathbf{W}_Q,\widetilde{\boldsymbol{\chi}}_{k}\mathbf{W}_K,\widetilde{\boldsymbol{\chi}}_{k}\mathbf{W}_V,\\
&\mathbf{a}_{k,t}=\mathrm{softmax}\!\left(\frac{\mathbf{Q}_{t}\mathbf{K}_{\Omega}^{\top}}{\sqrt{d_{attn}}}\right)\mathbf{V}_{\Omega},\qquad
t\in\mathcal{T}_f,\;\Omega=\mathcal{T}_h\cup\mathcal{T}_f.
\end{aligned}
\end{equation}

Unlike strict causal masking with $\Omega_{t}^{\mathrm{mask}}=\mathcal{T}_h\cup\{t'\in\mathcal{T}_f\mid t'\le t\}$, our forward-looking attention design allows each prediction step $t$ to incorporate information from the entire input window $\Omega$, including both the preceding context and anticipated disruption signals.

Finally, attention-enhanced representations over the future window are mapped to quantile outputs:

\begin{equation}
\begin{aligned}
&\hat{\mathbf{I}}_{k,t}^{\mathcal{Q}}=\mathrm{FNN}_{q}(\mathbf{a}_{k,t}),\qquad t\in\mathcal{T}_f,\;\mathcal{Q}=\{0.1,0.5,0.9\},\\
&\hat{I}_{k,t}^{(\tau)}=\left[\hat{\mathbf{I}}_{k,t}^{\mathcal{Q}}\right]_{\tau},\qquad \tau\in\mathcal{Q}.
\end{aligned}
\end{equation}

All quantile forecasts are supervised by quantile loss \citep{Wen2017multi}, while the median prediction $\hat{I}_{k,t}^{(0.5)}$ is used as the final point forecast for evaluation.

\subsection{Voltmeter-inspired bypass for latent demand inference}

Although the ammeter in Sec.~\ref{sec:ammeter} estimates disrupted passenger flow ($I$), the underlying driver, latent demand ($U$), is unobservable under disruption, but vital for understanding mobility flow dynamics. To infer $U$, we propose a voltmeter-inspired bypass parallel to the factual ammeter branch, where the context is replaced with a minimum-impedance reference.

\subsubsection{The voltmeter analogy}
The physical analogy is straightforward. In a circuit, a voltmeter can be built by connecting a known standard resistor in series with an ammeter and converting the measured current into voltage through Ohm's law. Similarly, \textit{OhmicFlow} introduces a parallel bypass that reuses the ammeter replica under normal conditions as a reference. Guided by Assumption~3, this reference is treated as a normalized resistor, i.e., $R=1$, so the bypass estimates latent demand rather than disrupted flow.
Specifically, this is implemented by replacing the anticipated disrupted travel time with normal travel time $\widetilde{\tau}_{k,t}$. For each OD pair $k$ and target time step $t\in\mathcal{T}_f$, we construct $\widetilde{\tau}_{k,t}$ from the historical travel times observed at the same hour of day and day of week, and take the 10th percentile as the reference value. This avoids the instability of a raw minimum while still keeping the reference close to a low-impedance normal status in which latent demand is sufficiently realized as passenger flow.

\subsubsection{Counterfactual forward mechanism}
Let $\mathcal{A}_{\theta}(\cdot)$ denote the ammeter network defined in Sec.~4.2. The bypass is implemented as a weight-sharing counterfactual pass of the ammeter network. It uses the same parameters $\theta$ as the factual pass, and these parameters are updated synchronously during training. The only difference is that the travel time input $\tau$ is replaced by its normal reference $\widetilde{\tau}$. Thus, the factual and counterfactual passes are given by

\begin{equation}
\begin{aligned}
&\hat{\mathbf{I}}_{k}^{\mathcal{T}_f,\mathcal{Q}} = \mathcal{A}_{\theta}\!\left(\mathbf{x}_{k}^{(h)},\mathbf{x}_{k}^{(f)}(\tau),\mathbf{x}_{k}^{(s)}\right), \\
&\hat{\mathbf{U}}_{k}^{\mathcal{T}_f,\mathcal{Q}} = \mathcal{A}_{\theta}\!\left(\mathbf{x}_{k}^{(h)},\mathbf{x}_{k}^{(f)}(\widetilde{\tau}),\mathbf{x}_{k}^{(s)}\right).
\end{aligned}
\end{equation}

The counterfactual estimate of latent demand is constrained in three ways. First, the normal travel time $\widetilde{\tau}$ is not an artificial value, but is constructed from historical observations under normal conditions. Replacing $\tau$ with $\widetilde{\tau}$ therefore moves the input toward a well-observed low-impedance condition rather than an unrealistic state outside of the data. Second, although latent demand $\hat{U}$ is not directly observed, it is still learned jointly with flow $\hat{I}$ and impedance $\hat{R}$ through the Ohmic consistency loss (details in Sec.~\ref{sec:loss}). If the inferred $\hat{U}$ becomes too large or too small, the resulting ratio $\hat{U}/\hat{R}$ would no longer be consistent with the predicted flow $\hat{I}$, producing a larger training loss and pushing the model back toward a physically reasonable range. Third, we impose a physical boundary $U\ge I$ during training (details in Sec.~\ref{sec:loss}), which prevents the inferred demand from being smaller than the passenger flow. Together, these mechanisms keep the estimated latent demand grounded in observed operating conditions, consistent with Ohm's law, and physically interpretable.

\subsection{PTC-informed dynamic impedance modeling}

After estimating $\hat{I}$ and $\hat{U}$, this section derives the dynamic OD impedance $\hat{R}$ to close the Ohmic loop. Although travel time $\tau$ acting as resistivity can be directly mapped to impedance, such a black-box shortcut blurs the effects of supply contraction and congestion effects during disruption, which is critical for robustness. To make these mechanisms explicit, we design a PTC-informed module that couples supply and congestion to infer physical impedance.

\subsubsection{The PTC analogy and formulation}
In PTC thermistors, increasing the current generates Joule heating, whose temperature rise approximately scales with the square of the current and then nonlinearly amplifies the base resistance. This base resistance is jointly shaped by PTC's normal-temperature conductivity, ambient states, and material properties. Similarly, in disrupted transit networks, base travel impedance is determined by supply conditions, weather, and static OD/station contexts. Congestion-induced service deterioration then acts like Joule heating, producing a nonlinear impedance increase above this base level.

\begin{figure}
    \centering
    \includegraphics[width=0.95\linewidth,height=0.95\textheight,keepaspectratio]{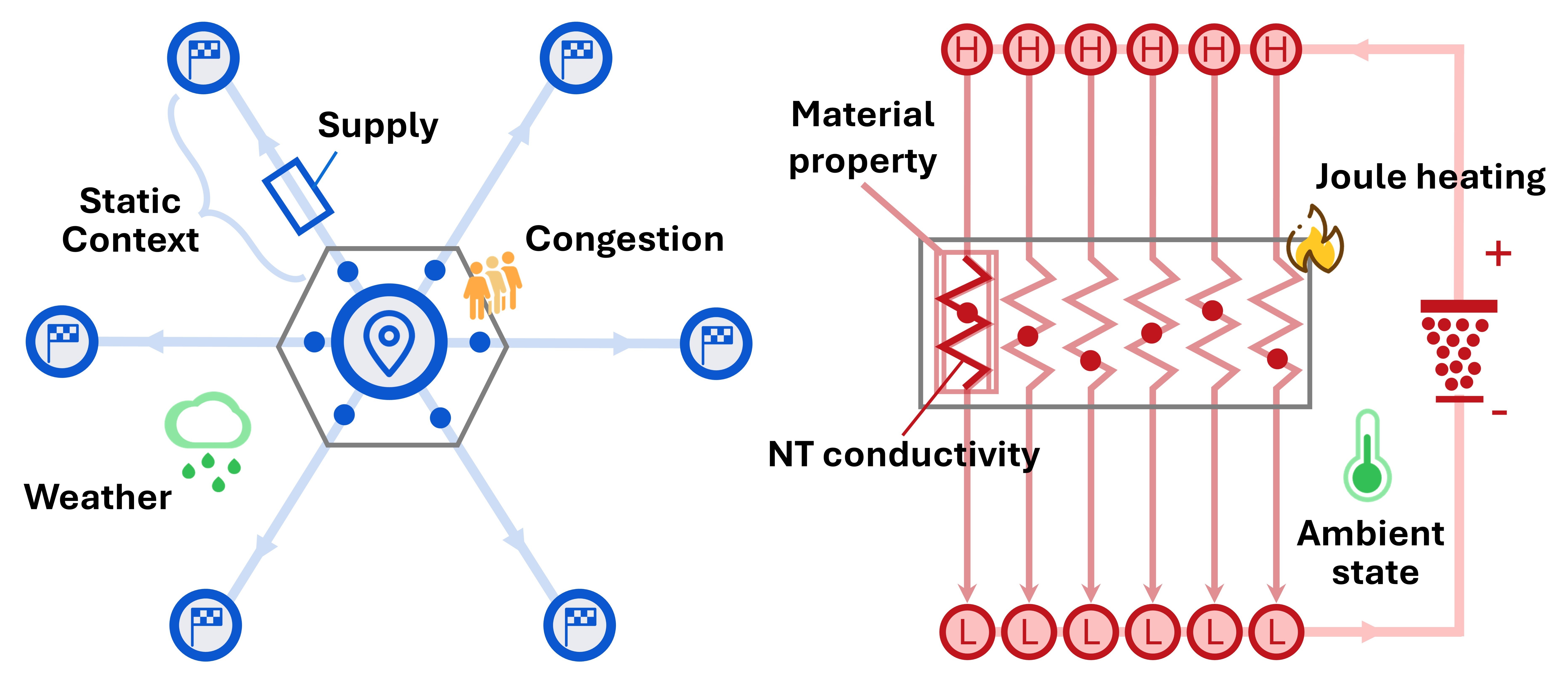}
    \vspace{-5pt}
    \captionof{figure}{The thermally coupled PTC analogy.}
    \label{fig:ptc}
\end{figure}

As shown in Fig.~\ref{fig:ptc}, congestion in transit systems is felt most directly at the origin station, where passengers first enter the disrupted system and encounter queues, platform crowding, and denied boarding. Essentially, the impedance of an OD pair depends not only on the OD-level conditions, but also on the aggregate inflow from all OD pairs sharing the same origin station. This resembles an indirectly heated PTC system, where a thermistor element is affected by both self-heating and heat transferred from nearby elements. We therefore model the spatial spillover of congestion using a thermally coupled PTC formulation:
\begin{equation}
\label{eq:ptc_spatial}
R_{k,t}=R^{b}_{k,t}\left[1+\alpha_{i,t}\left(\frac{I^{\mathrm{in}}_{i,t}}{\eta_{i,p}}\right)^2\right] + 1,\qquad k=(i,j).
\end{equation}
where $R^{b}_{k,t}$ denotes the base travel impedance, $I^{\mathrm{in}}_{i,t}$ the aggregated inflow at origin $i$, $\eta_{i,p}$ the historical peak inflow used as a station capacity proxy for scale normalization, and $\alpha_{i,t}$ is a time-varying station-level congestion sensitivity coefficient acting as PTC's temperature coefficient. Following the intuition of the Bureau of Public Roads (BPR) volume-delay function, we use the saturation ratio $I^{\mathrm{in}}_{i,t}/\eta_{i,p}$ instead of raw inflow, as congestion is better characterized by the ratio of inflow to capacity than by raw inflow volume.

In addition to spatial spillover, there may also exist temporal spillover under the quasi-steady assumption (i.e., Assumption 2), where disruption effects propagate across adjacent time steps. For each future step $t\in\mathcal{T}_f$, we define the local window and corresponding regularized impedance as
\begin{equation}
\label{eq:ptc_temporal}
\begin{aligned}
&\mathcal{N}_{t} = [t-\lfloor\Delta t/2\rfloor,\,t+\lfloor\Delta t/2\rfloor],\\
&\tilde{R}_{k,t} = \left(\text{avg}_{t'\in\mathcal{N}_{t}}R_{k,t'} + \text{max}_{t'\in\mathcal{N}_{t}}R_{k,t'}\right) / 2,
\end{aligned}
\end{equation}
where the averaging term stabilizes local fluctuations and the maximum term preserves the peak disruption intensity.

\subsubsection{Base impedance modeling}

The base impedance $R^b_{k,t}$ is primarily driven by the reduction in network suppy, which is difficult to measure directly because supply contraction can arise from multiple sources, such as station closure, train speed restriction, and frequency reduction. The average travel time $\tau_{k,t}$ can be used as the most direct observable proxy for supply conditions, though it is also affected by weather, congestion, temporal patterns, and static OD characteristics. Therefore, it is necessary to disentangle these confounding effects and extract supply-related signals. However, directly using raw $\tau_{k,t}$ is numerically problematic: (i) severe interruptions may yield $\tau_{k,t}\to\infty$, inducing instability and gradient singularity; and (ii) the normal travel time $\widetilde{\tau}_{k,t}$ has OD-specific heterogeneity, making the global assumption of $R_{k,t}=1$ for any OD pair under normal operation difficult to enforce. To address this, travel time is transformed into efficiency $e_{k,t}$ and relative efficiency $\Delta e_{k,t}$:
\begin{equation}
e_{k,t}=\omega\,\tau_{k,t}^{-1},\qquad
\Delta e_{k,t}=\max\!\left(\omega\widetilde{\tau}_{k,t}^{-1}-\omega\tau_{k,t}^{-1},\,0\right),
\end{equation}
where $\omega>0$ is a scaling constant. By construction, $e_{k,t}\in[0,a]$ is bounded, and $\Delta e_{k,t}=0$ under normal operation, providing a smooth and globally aligned reference signal.

In the ammeter branch, $\tau_{k,t}$ is also embedded through $(e_{k,t},\Delta e_{k,t})$ rather than raw time values. Unlike other variables, this pair is embedded through bias-free monotone mappings $\mathcal{M}^{\uparrow}(\cdot)$, a special case of $\phi(\cdot)$. The monotone constraint ensures that the embedding follows the numerical meaning of service quality: higher $e_{k,t}$ indicates better service efficiency, while larger $\Delta e_{k,t}$ indicates greater efficiency loss. This allows the numerical strength of these signals to be carried forward, so that higher efficiency or larger efficiency loss produces a stronger embedding magnitude instead of being weakened or reversed. The bias-free design further anchors the two zero boundaries: when $e_{k,t}=0$, the embedding remains at zero, indicating no effective service; when $\Delta e_{k,t}=0$, the embedding also remains at zero, representing no efficiency loss from normal operation. These boundary meanings are therefore not shifted by learned bias terms. Here, we extract the unified embeddings of $(e_{k,t},\Delta e_{k,t})$ and construct
\begin{equation}
\label{eq:eff_embed}
\boldsymbol{\epsilon}_{k,t}=\mathcal{M}^{\uparrow}_{e}(e_{k,t})\oslash\left(\mathcal{M}^{\uparrow}_{\Delta e}(\Delta e_{k,t})+\varepsilon\right),
\end{equation}
where $\oslash$ is element-wise division and $\varepsilon$ is a small constant. $\boldsymbol{\epsilon}_{k,t}$ reflects the comprehensive efficiency and satisfies two useful boundary properties: if $e_{k,t}=0$ (e.g., station shutdown/non-operation), then $\boldsymbol{\epsilon}_{k,t}=\mathbf{0}$ regardless of $\Delta e_{k,t}$; if $\Delta e_{k,t}=0$ with $e_{k,t}>0$ (normal-operation aligned), then $\|\boldsymbol{\epsilon}_{k,t}\|$ reaches a high and comparable scale across OD pairs, which can be further mapped to values close to 1 by the downstream $\tanh$ activation.

Given $\boldsymbol{\epsilon}_{k,t}$, we apply FiLM-style modulation \citep{Perez2018FiLM} to recover latent supply representation. The unified embeddings of $W_t$, $H_t$, $f_{i,p}$ and $f_{k,p}$ are used to filter out weather- and context-related components from $\boldsymbol{\epsilon}_{k,t}$. We then sum the ammeter's last states $\mathbf{a}_{k,t}$ by the origin to obtain the inflow representation $\mathbf{a}^{\mathrm{in}}_{i,t}$, thereby removing congestion-related components. The remaining component is then interpreted as the supply signal. Following the VSN design in the ammeter, we feed dynamic embeddings together with static embeddings, and then apply softplus gating to obtain: 
\begin{equation}
\label{eq:supply_film}
\begin{aligned}
&\boldsymbol{\gamma}_{k,t}=\operatorname{softplus}\!\left(\operatorname{VSN}\!\left([\phi(W_{k,t});\phi(H_{k,t});\mathbf{a}^{\mathrm{in}}_{i,t}],[\phi(f_{i,p});\phi(f_{k,p})]\right)\right),\\
&\mathbf{s}_{k,t}=\boldsymbol{\gamma}_{k,t}\odot\boldsymbol{\epsilon}_{k,t},\\
&S_{k,t}=\tanh\!\left(\mathcal{M}^{\uparrow}_{S}(\mathbf{s}_{k,t})\right) \in [0,1],
\end{aligned}
\end{equation}
where $\boldsymbol{\gamma}_{k,t}$ is the FiLM gating vector, $\mathbf{s}_{k,t}$ is the OD-level supply representation after gating, and $S_{k,t}$ is the scalar supply index. Here $S_{k,t}$ serves as an intermediate physical variable and is additionally used for auxiliary supervision (details in Sec.~\ref{sec:loss}). For base impedance estimation, we construct a second FiLM gate to remap the recovered supply state $\mathbf{s}_{k,t}$ under context semantics, because $R^b_{k,t}$ should be jointly determined by supply, weather and other context. Compared with $\boldsymbol{\gamma}_{k,t}$, the new gate $\boldsymbol{\gamma}^{b}_{k,t}$ removes the origin-inflow representation $\mathbf{a}^{\mathrm{in}}_{i,t}$ to ensure that the congestion-induced impedance is separated from the base impedance. The base-impedance mapping is then defined as
\begin{equation}
\label{eq:base_impedance}
\begin{aligned}
&R^{b}_{k,t}= R^{\max}\left(1-\tanh\!\left(\mathcal{M}^{\uparrow}_{b}(\boldsymbol{\gamma}^{b}_{k,t}\odot\mathbf{s}_{k,t})\right)\right) \in[0,R^{\max}],
\end{aligned}
\end{equation}
where $R^{\max}$ denotes the theoretical upper bound of base impedance, enforced to ensure numerical stability during model training. Overall, this design yields a monotone chain $e_{k,t}\uparrow, \Delta e_{k,t}\downarrow\Rightarrow\|\boldsymbol{\epsilon}_{k,t}\|\uparrow\Rightarrow\|\mathbf{s}_{k,t}\|\uparrow\Rightarrow S_{k,t}\uparrow, R^b_{k,t}\downarrow$, and enforces boundary-consistent behavior: normal states map to $R^{b}_{k,t}=0$ and $R_{k,t}=1$.

\subsubsection{Congestion term modeling}

To evaluate the congestion amplification term $C_{i,t}=\alpha_{i,t}({I^{\mathrm{in}}_{i,t}}/{\eta_{i,p}})^2$ in the PTC formulation, we need to estimate $I^{\mathrm{in}}_{i,t}$ and $\alpha_{i,t}$, while $\eta_{i,p}$ is treated as a fixed hyperparameter determined in the data processing stage.

For station inflow, we distinguish training and inference. During training, we directly read the ground-truth OD flow $I_{k,t}$ from each batch input and aggregate them by origin $I^{\mathrm{in}}_{i,t}=\sum_{j \in V_p} I_{k=(i,j),t}$,
which avoids error accumulation from recursive training. During inference, the same aggregation is performed on the median ammeter output $\hat{I}^{(0.5)}_{k,t}$, since the ground-truth flow is unavailable at test time.

The congestion sensitivity is modeled as a dynamic rather than static quantity, since station crowding depends jointly on intrinsic station features and dynamic transit supply. We first construct a station-level supply representation,
\begin{equation}
\mathbf{s}^{\mathrm{in}}_{i,t}=\frac{\sum_{j \in V_p} I_{k=(i,j),t}\,\mathbf{s}_{k=(i,j),t}}{I^{\mathrm{in}}_{i,t}},
\end{equation}
which summarizes the average supply state for OD pairs with the same origin. We then define the station-level congestion sensitivity as
\begin{equation}
\label{eq:congestion_film}
\begin{aligned}
&\boldsymbol{\gamma}_{i,t}=\operatorname{softplus}\!\left(\phi(f_{i,p})\right)\otimes\mathbf{1}_{t},\\
&\alpha_{i,t}=\alpha^{\max}\left(1-\tanh\!\left(\mathcal{M}^{\uparrow}_{\alpha}\!\left(\boldsymbol{\gamma}_{i,t}\odot\mathbf{s}^{\mathrm{in}}_{i,t}\right)\right)\right),
\end{aligned}
\end{equation}
where $\boldsymbol{\gamma}_{i,t}$ is the FiLM gate obtained from the station embedding and expanded over time. 
This ensures $\alpha_{i,t}$ is bounded by $\alpha^{\max}$, monotone increasing with supply contraction, and reflective of both static features and dynamic supply.

Taken together, $\alpha_{i,t}$, $I^{\mathrm{in}}_{i,t}$, and $\eta_{i,p}$ provide a complete and interpretable description of congestion-induced Joule heating. Their combination yields the final nonlinear amplification term $C_{i,t}$, and we further impose an auxiliary supervision signal on this quantity in Sec.~\ref{sec:loss} to strengthen physical consistency during training.

\subsection{Training strategies}

\subsubsection{Physics-constrained multi-objective loss}\label{sec:loss}

The total loss comprises three components: predictive fidelity ($\mathcal{L}_{\mathrm{data}}$), Ohmic consistency ($\mathcal{L}_{\mathrm{ohm}}$), and soft physical supervision ($\mathcal{L}_{\mathrm{soft}}$). Let $\mathcal{B}$ and $\mathcal{B}_{\mathrm{in}}$ denote the mini-batch of sampled OD pairs and their unique origins. Using the quantile penalty $\rho_q(x)=\max(qx,(q-1)x)$, the predictive fidelity term is given as

\begin{equation}
\begin{aligned}
&\mathcal{L}_{\mathrm{od}}=\frac{1}{|\mathcal{B}|\,|\mathcal{T}_f|\,|\mathcal{Q}|}\sum_{k\in\mathcal{B}}\sum_{t\in\mathcal{T}_f}\sum_{q\in\mathcal{Q}}\rho_q\!\left(I_{k,t}-\hat{I}_{k,t}^{(q)}\right),\\
&\mathcal{L}_{\mathrm{in}}=\frac{1}{|\mathcal{B}_{\mathrm{in}}|\,|\mathcal{T}_f|\,|\mathcal{Q}|}\sum_{i\in\mathcal{B}_{\mathrm{in}}}\sum_{t\in\mathcal{T}_f}\sum_{q\in\mathcal{Q}}\rho_q\!\left(I_{i,t}^{\mathrm{in}}-\hat{I}_{i,t}^{\mathrm{in},(q)}\right),\\
&\mathcal{L}_{\mathrm{data}}=\mathcal{L}_{\mathrm{od}}+\lambda_{\mathrm{in}}\mathcal{L}_{\mathrm{in}}.
\end{aligned}
\end{equation}
where \(\mathcal{L}_{\mathrm{od}}\) is the loss associated with OD passenger flow prediction, while \(\mathcal{L}_{\mathrm{in}}\) adds the same-origin inflow constraint inspired by KCL and prior flow-conservation regularization studies \citep{Zhang2024Physics}; together they form the predictive fidelity term.

For Ohmic consistency, the ground-truth OD flow is again used to supervise the ratio between inferred demand and estimated impedance. We still retain the quantile loss and define Ohmic consistency as
\begin{equation}
\mathcal{L}_{\mathrm{ohm}}=\frac{1}{|\mathcal{B}|\,|\mathcal{T}_f|\,|\mathcal{Q}|}\sum_{k\in\mathcal{B}}\sum_{t\in\mathcal{T}_f}\sum_{q\in\mathcal{Q}}\left|I_{k,t}-\frac{\hat{U}_{k,t}^{(q)}}{\tilde{R}_{k,t}}\right|.
\end{equation}

Since $\mathcal{L}_{\mathrm{ohm}}$ essentially supervises flow through the demand-impedance coupling under the Ohmic constraint, it is weighted equally with \(\mathcal{L}_{\mathrm{od}}\) to keep the two primary supervision channels balanced.

To improve the robustness of the Ohmic constraint, we regularize physical latent states by introducing a soft supervision term \(\mathcal{L}_{\mathrm{soft}}\), which injects three directional and boundary priors:
\begin{equation}
\begin{aligned}
&\mathcal{L}_{U\geq I}=\frac{1}{|\mathcal{B}|\,|\mathcal{T}_f|\,|\mathcal{Q}|}\sum_{k\in\mathcal{B}}\sum_{t\in\mathcal{T}_f}\sum_{q\in\mathcal{Q}}\operatorname{softplus}\!\left(\kappa\,\frac{\hat{I}_{k,t}^{(q)}-\hat{U}_{k,t}^{(q)}}{\hat{I}_{k,t}^{(q)}+\varepsilon}\right),\\
&\mathcal{L}_{\mathrm{sup}}=1-\operatorname{Corr}\!\left(\hat{S}_{k,t},\frac{e_{k,t}}{e_{k,t}+\Delta e_{k,t}+\varepsilon}\right)_{(k,t)\in\Omega_{\mathrm{wc}}},\\
&\mathcal{L}_{\mathrm{con}}=1-\operatorname{Corr}\!\left(\hat{C}_{i,t},\frac{1}{|V_p|}\sum_{j \in V_p}\frac{\tau_{(i,j),t}-\tau_{(j,i),t}}{\tau_{(i,j),t}+\tau_{(j,i),t}}\right)_{(i,t)\in\mathcal{B}_{\mathrm{in}}\times\mathcal{T}_f},\\
&\mathcal{L}_{\mathrm{soft}}=\lambda_{U}\mathcal{L}_{U\geq I}+\lambda_{S}\mathcal{L}_{\mathrm{sup}}+\lambda_{C}\mathcal{L}_{\mathrm{con}}.
\end{aligned}
\end{equation}

Specifically, $\mathcal{L}_{U\geq I}$ penalizes nonphysical violations where passenger flow exceeds latent demand ($\kappa$ scales sharpness, $\varepsilon$ prevents zero-division). $\mathcal{L}_{\mathrm{sup}}$ aligns the supply capacity index $\hat{S}_{k,t}$ with relative efficiency under weak congestion ($\Omega_{\mathrm{wc}}=\{(k,t) \in\mathcal{B}\times\mathcal{T}_f:\hat{C}_{i,t}<0.1\}$), where efficiency is mainly governed by supply contraction. $\mathcal{L}_{\mathrm{con}}$ provides weak supervision for the congestion term $\hat{C}_{i,t}$, which is otherwise unobserved and could be confused with supply-induced impedance. We use bidirectional travel time asymmetry as an observable proxy: for an OD pair $i\rightarrow j$ and its reverse pair $j\rightarrow i$, the in-vehicle travel time is expected to be similar and destination-side congestion is usually negligible. Therefore, the remaining asymmetry mainly reflects congestion on the origin-side, such as platform crowding, queuing, and denied boarding, and is used to guide the estimation of $\hat{C}_{i,t}$.

Putting the three parts together, the full training objective is
\begin{equation}
\mathcal{L}=\mathcal{L}_{\mathrm{data}}+\mathcal{L}_{\mathrm{ohm}}+\mathcal{L}_{\mathrm{soft}},
\end{equation}

\subsubsection{Origin-centric batching with FIFO aggregation}

Estimating the full OD matrix in each batch is computationally expensive, while naive random shuffling and sampling of OD pairs breaks the station-level aggregations (e.g., $I^{\mathrm{in}}_{i,t}$, $\mathbf{s}^{\mathrm{in}}_{i,t}$) required by \textit{OhmicFlow}. Thus, we group all OD pairs that share the same origin into a single batch, enabling direct aggregation and reducing spatial sampling overhead through highly overlapped $\mathcal{N}_{p,k}$ across OD pairs.

However, since each batch covers only a subset of origins, its loss may be biased toward the spatial and demand patterns of the selected origin groups. To reduce such batch-specific bias, we use a first-in-first-out (FIFO) loss aggregator that buffers recent batch losses from different origin groups and optimizes them jointly. Let \(\mathcal{F}^{\mathrm{buf}}_m\) denote the queue after the \(m\)-th batch, $M$ the FIFO capacity, and \(\mathcal{L}_m\) the current batch loss. The update is written as
\begin{equation}
\begin{aligned}
&\mathcal{F}^{\mathrm{buf}}_m = \mathrm{FIFO}(\mathcal{F}^{\mathrm{buf}}_{m-1},\mathcal{L}_m, M),\\
&\mathcal{L}^{\mathrm{agg}}_m = \frac{1}{|\mathcal{F}^{\mathrm{buf}}_m|}\sum_{\mathcal{L}\in\mathcal{F}^{\mathrm{buf}}_m}\mathcal{L}.
\end{aligned}
\end{equation}
where \(\mathrm{FIFO}(\cdot)\) appends the new loss and discards the oldest entry when the queue is full. This aggregator recovers more diverse gradients from heterogeneous origins through delayed joint optimization. To summarize the process of training and inference more intuitively, we formalize the procedure in the pseudo-code algorithm shown in Algorithm~\ref{alg:most}.

\begin{algorithm}[!t]
\caption{Training and inference of \textit{OhmicFlow}}
\label{alg:most}
\begin{algorithmic}[1]
\Require OD neighbors \(\mathcal{N}_{p,k}\), Quantile set \(\mathcal{Q}\), FIFO capacity \(M\), Raw inputs \(X=\{\mathbf{x}_{k}^{(h)},\mathbf{x}_{k}^{(f)},\mathbf{x}_{k}^{(s)}\}\)
\Ensure Trained parameters \(\theta\); inference outputs \(\hat{\mathbf{I}}_{k}^{\mathcal{Q}}\), \(\hat{\mathbf{U}}_{k}^{\mathcal{Q}}\) and \(\tilde{R}_{k,t}\) during $\mathcal{T}_f$, together with \(R^b_{k,t}\), \(\alpha_{i,t}\), and \(S_{k,t}\)
\State Initialize \(\theta\) and FIFO buffer \(\mathcal{F}^{\mathrm{buf}}_0\)
\For{each epoch}
    \State \textbf{// Step1: Origin-centric Batching}
    \State Sample an origin subset \(\mathcal{O}_m \subset V_p\)
    \State Extract target samples \(\mathcal{B}_m^{\mathrm{tar}}\) by collecting all ODs originating from \(\mathcal{O}_m\)
    \State Extract neighbor samples \(\mathcal{B}_m^{\mathrm{nbr}}(k)\) from \(\mathcal{N}_{p,k}\) for each OD pair \(k\)
    \State Build batch \(\mathcal{B}_m\) by combining \(\mathcal{B}_m^{\mathrm{tar}}\) with all \(\mathcal{B}_m^{\mathrm{nbr}}(k)\) \Comment{targets are supervised; neighbors provide context}
    \State \textbf{// Step2: Ammeter Predicts Disrupted Flow}
    \State Embed $\mathbf{x}_{k}^{(h)}, \mathbf{x}_{k}^{(f)}, \mathbf{x}_{k}^{(s)}$ via $\phi(\cdot)$ for \(k\in\mathcal{B}_m\) (Specifically, for $e_{k,t}$ and $\Delta e_{k,t}$, $\phi(\cdot) = \mathcal{M}^{\uparrow}(\cdot)$)  
    \State Predict \(\hat{\mathbf{I}}_{k}^{\mathcal{Q}}\) with \(\mathcal{A}_{\theta}(\cdot)\) for \(k\in\mathcal{B}_m^{\mathrm{tar}}\)  \Comment{factual forward pass}
    \State \textbf{// Step3: Voltmeter Infers Latent Demand}
    \State Replace anticipated disrupted travel time \({\tau}_{k,t}\) with normal reference \(\widetilde{\tau}_{k,t}\) for \(k\in\mathcal{B}_m\) and \(t\in\mathcal{T}_f\)
    \State Predict \(\hat{\mathbf{U}}_{k}^{\mathcal{Q}}\) with the shared \(\mathcal{A}_{\theta}(\cdot)\) and modified inputs for \(k\in\mathcal{B}_m^{\mathrm{tar}}\)  \Comment{counterfactual forward pass}
    \State \textbf{// Step4: PTC Models Dynamic Impedance}
    \State Fetch $\phi(\mathbf{x}_{k}^{(f)})$ and $\phi(\mathbf{x}_{k}^{(s)})$ from Ammeter for \(k\in\mathcal{B}_m^{\mathrm{tar}}\),
    \State Consturct comprehensive efficiency states: \(\boldsymbol{\epsilon}_{k,t} \leftarrow \mathcal{M}^{\uparrow}(e_{k,t}), \mathcal{M}^{\uparrow}(\Delta e_{k,t})\) (Eq.~\eqref{eq:eff_embed})
    \State Modulate hidden OD supply states: \(\mathbf{s}_{k,t} \leftarrow \boldsymbol{\epsilon}_{k,t}, \phi(\mathbf{x}_{k}^{(f)}), \phi(\mathbf{x}_{k}^{(s)})\); \(S_{k,t} \leftarrow \mathbf{s}_{k,t}\) (Eq.~\eqref{eq:supply_film})
    \State Modulate OD base impedance: \(R^{b}_{k,t} \leftarrow \mathbf{s}_{k,t}, \phi(\mathbf{x}_{k}^{(f)}), \phi(\mathbf{x}_{k}^{(s)})\) (Eq.~\eqref{eq:base_impedance})
    \If{Training}
        \State Aggregate \(I^{\mathrm{in}}_{i,t}\) and \(\mathbf{s}^{\mathrm{in}}_{i,t}\) from ground-truth OD flows \(I_{i,t}\)
    \Else
        \State Aggregate \(I^{\mathrm{in}}_{i,t}\) and \(\mathbf{s}^{\mathrm{in}}_{i,t}\) from median forecasts \(\hat{I}^{(0.5)}_{k,t}\) of Ammeter
    \EndIf
    \State Modulate station-level congestion sensitivity: \(\alpha_{i,t} \leftarrow \mathbf{s}^{\mathrm{in}}_{i,t}, \phi(\mathbf{x}_{k}^{(s)})\) (Eq.~\eqref{eq:congestion_film})
    \State Compute OD impedance \(R_{k,t}\) according to PTC formulation (Eq.~\eqref{eq:ptc_spatial}) \Comment{control spatial spillover}
    \State Compute final regularized impedance\(\tilde{R}_{k,t}\) (Eq.~\eqref{eq:ptc_temporal}) \Comment{control temporal spillover}
    \State \textbf{// Step5: FIFO Loss Optimization or Output Return}
    \If{Training}
        \State Compute \(\mathcal{L}_m=\mathcal{L}_{\mathrm{data}}+\mathcal{L}_{\mathrm{ohm}}+\mathcal{L}_{\mathrm{soft}}\)
        \State Push \(\mathcal{L}_m\) into the FIFO buffer: \(\mathcal{F}^{\mathrm{buf}}_m \leftarrow \mathrm{FIFO}(\mathcal{F}^{\mathrm{buf}}_{m-1},\mathcal{L}_m,M)\)
        \State Minimize \(\mathcal{L}^{\mathrm{agg}}_m=\frac{1}{|\mathcal{F}^{\mathrm{buf}}_m|}\sum_{\mathcal{L}\in\mathcal{F}^{\mathrm{buf}}_m}\mathcal{L}\) and update \(\theta\)
    \Else
        \State Return \(\hat{\mathbf{I}}_{k}^{\mathcal{Q}}\), \(\hat{\mathbf{U}}_{k}^{\mathcal{Q}}\), \(\tilde{R}_{k,t}\), \(R^{b}_{k,t}\), \(\alpha_{i,t}\), and \(S_{k,t}\)
    \EndIf
\EndFor
\end{algorithmic}
\end{algorithm}

\section{Case study}\label{sec:case}
\subsection{Dataset description}

To demonstrate our proposed methodology, a case study is conducted on Shenzhen Metro, an urban rail transit system that serves as the backbone of daily urban mobility in Shenzhen, a coastal megacity located in southern China. Shenzhen Metro carries around 7.43 million passengers per day and accounts for more than 70\% of the city's public transport trips \citep{Shenzhen2024Shenzhen}. The high ridership, combined with Shenzhen's location along a typhoon-prone coastline, makes the system especially vulnerable to EWE-induced disruptions and therefore an appropriate testbed for \textit{OhmicFlow}.

Our dataset combines Automatic Fare Collection (AFC) records from Shenzhen Metro and hourly weather data in Shenzhen during a 10-year study period from 2014 to 2023. Based on Shenzhen's annual climate bulletins, we identify 17 disruptive typhoon and rainstorm events that produced noticeable mobility impacts, as summarized in Table~\ref{tab:shenzhen_events}. While representing only a subset of EWEs in our study period, these 17 events provide an adequate and diverse sample to validate the efficacy of \textit{OhmicFlow}.

\begin{table*}
\centering
\small
\caption{Disruptive weather events in Shenzhen during the study period}
\label{tab:shenzhen_events}
\begin{tabular}{p{0.7cm}p{3.5cm}p{5.6cm}p{1.5cm}p{2.5cm}}
\hline
\textbf{No.} & \textbf{Period} & \textbf{Event} & \textbf{Max wind} & \textbf{Max daily rainfall} \\
\hline
1 & 2014/05/08--2014/05/09 & Extreme rainstorm triggered by cold air & 7 m/s & 313 mm \\
2 & 2014/05/11--2014/05/11 & Extreme rainstorm triggered by cold air & 8 m/s & 443 mm \\
3 & 2016/04/10--2016/04/10 & Rainstorm triggered by severe convection & 17 m/s & 186 mm \\
4 & 2016/08/01--2016/08/03 & Typhoon NIDA (Level 14) & 42 m/s & 166 mm \\
5 & 2017/06/12--2017/06/13 & Typhoon MERBOK (Level 10) & 25 m/s & 162 mm \\
6 & 2017/09/03--2017/09/04 & Typhoon MAWAR (Level 10) & 25 m/s & 150 mm \\
7 & 2019/07/02--2019/07/03 & Typhoon MUN (Level 8) & 18 m/s & 40 mm \\
8 & 2019/07/31--2019/08/02 & Typhoon WIPHA (Level 9) & 23 m/s & 92 mm \\
9 & 2019/09/02--2019/09/04 & Typhoon KAJIKI (Level 8) & 18 m/s & 74 mm \\
10 & 2021/10/08--2021/10/10 & Typhoon LIONROCK (Level 8) & 20 m/s & 177 mm \\
11 & 2021/10/12--2021/10/14 & Typhoon KOMPASU (Level 12) & 35 m/s & 52.5 mm \\
12 & 2022/08/08--2022/08/10 & Typhoon MULAN (Level 9) & 23 m/s & 126 mm \\
13 & 2023/03/25--2023/03/26 & Extreme rainstorm triggered by cold air & 23 m/s & 161 mm \\
14 & 2023/07/16--2023/07/18 & Typhoon TALIM (Level 13) & 40 m/s & 95 mm \\
15 & 2023/09/01--2023/09/02 & Typhoon SAOLA (Level 16) & 62 m/s & 191 mm \\
16 & 2023/09/07--2023/09/08 & Typhoon HAIKUI (Level 16) & 52 m/s & 496 mm \\
17 & 2023/10/08--2023/10/09 & Typhoon KOINU (Level 16) & 55 m/s & 252 mm \\
\hline
\end{tabular}
\end{table*}

Notably, during the study period, Shenzhen Metro underwent 10 network expansions, which altered the network topology and therefore affected both station-level and OD-level structural context. Therefore, key network context features, such as station centrality and OD redundancy indicators, stay static during any specific event, but can vary over time across events. In addition, the weather data contain hourly rainfall and wind conditions as well as the weather signal status released by the Shenzhen meteorological authority. These weather variables are encoded as ordered discrete states, where larger values indicate more severe weather conditions. Dynamic temporal context variables include the month, day of week, and hour of day. The AFC records provide hourly OD passenger flow and average travel time for every OD pair. These data sources are further processed into the model inputs \(\mathbf{x}_{k}^{(h)},\mathbf{x}_{k}^{(f)},\mathbf{x}_{k}^{(s)}\).

\subsection{Experiment settings}

To evaluate \textit{OhmicFlow}'s adaptability to diverse weather conditions, the 17 events are chronologically partitioned into three progressive training sets (events 1-5, 6-10, and 11-14). For each set, the remaining out-of-sample events are designated as the test set. Notably, the final three events (15-17) are strictly withheld across all splits as a universal test set; representing the most recent and severe disruptions, they provide a rigorous benchmark for the model's capacity to generalize to unseen shocks.

During both training and testing, each epoch iterates over all selected extreme days. For each day, we randomly sample six distinct pairs of historical and future windows, $(\mathcal{T}_h, \mathcal{T}_f)$. A sample is valid when its $\mathcal{T}_h$ or $\mathcal{T}_f$ have at least 50\% of the time window in a day affected by an EWE. In this way, each valid sample contains a substantial disruption signal, either as part of the observed history for model inputs or as part of the future period to be predicted. The valid samples are then fed into Algorithm~\ref{alg:most}.

For the standard model configuration, we establish the key hyperparameters as: $\Delta T=12$, $\Delta t=3$, $d=64$,  $M=200$, $R^{\max}=4$, $\alpha^{max}=10$, and $|\mathcal{O}_m|=16$. The coefficients for the physical constraints are fixed to $\lambda_U=\lambda_S=\lambda_C=10$, while $\lambda_{\mathrm{in}}$ is independently determined per training split through validation. These parameter settings yielded favorable performance during preliminary studies.

For evaluation, we consider two dimensions: accuracy and uncertainty. Accuracy metrics include mean absolute error (MAE), root mean squared error (RMSE), and symmetric mean absolute percentage error (sMAPE):

\begin{equation}
\begin{aligned}
&\mathrm{MAE}=\frac{1}{|\Omega|}\sum_{(k,t)\in\Omega}\left|I_{k,t}-\hat{I}_{k,t}^{(0.5)}\right|,\\
&\mathrm{RMSE}=\sqrt{\frac{1}{|\Omega|}\sum_{(k,t)\in\Omega}\left(I_{k,t}-\hat{I}_{k,t}^{(0.5)}\right)^2},\\
&\mathrm{sMAPE}=\frac{1}{|\Omega|}\sum_{(k,t)\in\Omega}\frac{2\left|I_{k,t}-\hat{I}_{k,t}^{(0.5)}\right|}{|I_{k,t}|+\left|\hat{I}_{k,t}^{(0.5)}\right|+\varepsilon},
\end{aligned}
\end{equation}
where $\Omega$ is the set of evaluated OD-time pairs. For uncertainty quantification, we compute the prediction interval coverage probability (PICP) and mean prediction interval width (MPIW) as
\begin{equation}
\begin{aligned}
&\mathrm{PICP}=\frac{1}{|\Omega|}\sum_{(k,t)\in\Omega}\mathbb{I}\!\left(\hat{I}_{k,t}^{(0.1)}\le I_{k,t}\le\hat{I}_{k,t}^{(0.9)}\right),\\
&\mathrm{MPIW}=\frac{1}{|\Omega|}\sum_{(k,t)\in\Omega}\left(\hat{I}_{k,t}^{(0.9)}-\hat{I}_{k,t}^{(0.1)}\right),
\end{aligned}
\end{equation}
where $\mathbb{I}(\cdot)$ is an indicator function that returns 1 when the observed flow falls within the 10\%--90\% prediction interval and 0 otherwise. Together, these metrics quantify whether \textit{OhmicFlow} simultaneously achieves high point accuracy and well-calibrated uncertainty intervals.

\subsection{Baseline models} \label{sec:baselines}
Model comparison analysis is conducted against several representative deep learning baselines, both to evaluate the overall performance of \textit{OhmicFlow} and to validate the extensibility of its Ohmic constraint across these diverse architectures. These models are categorized into four groups based on their architectural paradigms.

First, two baselines are pure temporal models:
\begin{itemize}[noitemsep]
    \item \textbf{LSTM}\citep{Hochreiter1997Long} overcomes the problem of vanishing gradients by using specialized memory cells and gating mechanisms to learn and store long-term temporal dependencies. In our implementation, the LSTM baseline serves as an ablated version of TFT-GAT, omitting the GAT module, future-aware attention, and the GRN/VSN feature-enhancement blocks.
    \item \textbf{Transformer} \citep{Vaswani2017Attention} relies entirely on self-attention mechanisms to
    model long-range temporal dependencies, dispensing with recurrence and convolutions. Architecturally, the Transformer baseline here is equivalent to the TFT-GAT framework stripped of its LSTM, GAT, and GRN/VSN components.
\end{itemize}

Second, three baselines are convolution-based spatiotemporal models:
\begin{itemize}[noitemsep]
    \item \textbf{STGCN} \citep{Yu2018Spatio} integrates spatial graph convolution layers with 1D temporal convolution, enabling efficient extraction of both spatial topological features and temporal dynamic behaviors. Since the original STGCN only considers historical sequences, we pass the future-window embeddings through a single-step MLP to perform an additive correction on the backbone output, ensuring a fair comparison under the same input scope.
    \item \textbf{DCRNN} \citep{Li2018Diffusion} integrates diffusion convolutions based on bidirectional random walks to capture spatial topology, while utilizing a sequence-to-sequence framework with recurrent units for autoregressive temporal decoding. Following the original encoder-decoder logic, we feed the single-step future-window embeddings into the DCGRUCell to facilitate recursive decoding.
    \item \textbf{Graph WaveNet} \citep{Wu2019Graph} combines dilated temporal convolutions with adaptive graph learning for long-horizon forecasting. Similar to STGCN, an MLP-based additive correction is applied to incorporate future-window embeddings into its history-only architecture.
\end{itemize}

Third, two baselines are attention-based spatiotemporal models:
\begin{itemize}[noitemsep]
    \item \textbf{GMAN} \citep{Zheng2020GMAN} employs an attention-based encoder-decoder to model spatio-temporal dynamics, utilizing a transform attention mechanism to bridge historical and future steps and mitigate error propagation in long-term forecasting. To mitigate the computational overhead of full-graph dense attention, we here execute sparse spatial attention within our local graph $\mathcal{N}_{p,k}$ to maintain a consistent complexity. Future embeddings are used as queries to align historical memory as in the original mechanism, with causal masking retained.
    \item \textbf{STTN} \citep{Xu2021Spatial} captures complex, time-varying spatio-temporal patterns for long-term forecasting by coupling a spatial transformer for real-time directed spatial dependencies with a temporal transformer for long-range bidirectional modeling. We also replace dense spatial attention with sparse attention over $\mathcal{N}_{p,k}$. Following the original logic, future embeddings are concatenated with encoder outputs before decoder self-attention, while causal masking is preserved.
\end{itemize}

Finally, two baselines are physics-informed models:
\begin{itemize}[noitemsep]
    \item \textbf{PI-MPN} \citep{Wu2025Physics} integrates a neural diffusion network, treating human movement as a diffusion process where the marginal sums of the predicted OD matrix exactly matched diffusion-governed boundary flows. Due to the lack of granular point-of-interest (POI) data, our implementation substitutes the original gravity-model-based adaptive graph augmentation with local attention operations defined by $\mathcal{N}_{p,k}$.
    \item \textbf{PAG-STAN} \citep{Zhang2024Physics} embeds the physical quantity conservation between OD flows and station inflows into a masked physics-guided loss function to enhance interpretability, while utilizing dynamic compression to mitigate typical data sparsity. In our setting, we remove its real-time OD estimation module because real-time OD reconstruction is outside the scope of this study.
\end{itemize}

To ensure a fair and controlled comparison, all baselines are evaluated under unified settings. First, all models use the same input embedding pipeline as \textit{OhmicFlow} before entering their own temporal/spatial backbones. This standardization ensures a consistent starting point and facilitates the validation of \textit{OhmicFlow}'s extensibility across these baseline architectures. Second, to guarantee uncertainty evaluation, all baselines terminate with a quantile output head. Third, for architectures originally ignoring static context, static features are broadcast across the temporal dimension as exogenous covariates. Finally, all baselines are trained with the same data splits, optimization schedule, and number of epochs as \textit{OhmicFlow}; however, baseline optimization uses only $\mathcal{L}_{\mathrm{od}}$ as the training objective when comparing overall performance to reflect their native learning capacities.

\section{Results}\label{sec:results}
\subsection{Overall performance}

\begin{table}[t]
  \centering
  \caption{Performance comparison of baseline models across varying event periods.}
  \label{tab:performance_comparison}
  \resizebox{\textwidth}{!}{
    \begin{tabular}{lccccccccccccccc}
      \toprule
      \multirow{2}{*}{Model} & \multicolumn{5}{c}{$\text{Event}_{1-5}$} & \multicolumn{5}{c}{$\text{Event}_{6-10}$} & \multicolumn{5}{c}{$\text{Event}_{11-14}$} \\
      \cmidrule(lr){2-6} \cmidrule(lr){7-11} \cmidrule(lr){12-16}
      & MAE & RMSE & sMAPE & MPIW & PICP & MAE & RMSE & sMAPE & MPIW & PICP & MAE & RMSE & sMAPE & MPIW & PICP \\
      \midrule
      LSTM & 7.41 & 10.78 & 51.09 & 29.03 & 87.31 & 10.04 & 13.16 & 64.52 & 27.85 & 61.84 & 4.61 & 6.96 & 38.84 & 13.92 & 81.55 \\
      Transformer & \underline{5.25} & \underline{8.04} & \underline{43.07} & \underline{22.84} & \textbf{95.10} & 6.88 & 9.38 & 51.97 & 19.54 & 77.63 & 4.03 & 6.17 & 32.64 & 16.46 & \underline{94.76} \\
      STGCN & 8.18 & 11.35 & 58.42 & 23.60 & 71.49 & 6.86 & 9.09 & 56.22 & 19.15 & 68.01 & 4.73 & 6.68 & 43.71 & 17.14 & 76.46 \\
      DCRNN & 6.36 & 9.04 & 51.99 & 33.31 & 92.80 & 4.16 & 6.20 & 37.89 & 20.76 & \textbf{97.23} & 4.02 & 6.33 & 34.13 & 20.88 & 94.67 \\
      GWN & 8.77 & 11.77 & 62.48 & 25.35 & 77.96 & 5.01 & 7.08 & 45.26 & 15.04 & 77.98 & 4.45 & 6.49 & 40.30 & 16.09 & 85.41 \\
      GMAN & 12.27 & 17.87 & 65.63 & 70.66 & 84.06 & 5.16 & 7.47 & 43.62 & 22.99 & 84.82 & 4.86 & 7.04 & 41.79 & 20.26 & 84.67 \\
      STTN & 7.89 & 10.92 & 55.76 & 23.71 & 76.64 & 6.03 & 8.54 & 46.48 & 18.27 & 75.83 & 4.90 & 7.24 & 39.27 & 15.01 & 82.21 \\
      PI-MPN & 12.65 & 16.93 & 70.24 & 40.71 & 54.34 & 5.93 & 8.52 & 46.65 & 22.85 & 67.69 & 6.15 & 8.74 & 47.98 & 16.78 & 62.25 \\
      PAG-STAN & 6.95 & 9.74 & 52.57 & 25.53 & 78.73 & \underline{3.79} & \underline{5.63} & \underline{34.91} & \underline{15.04} & 87.80 & \underline{3.64} & \underline{5.65} & \underline{31.88} & \textbf{12.78} & 88.09 \\
      \textbf{OhmicFlow (ours)} & \textbf{3.22} & \textbf{4.97} & \textbf{29.37} & \textbf{13.40} & \underline{93.62} & \textbf{3.11} & \textbf{4.89} & \textbf{26.93} & \textbf{12.12} & \underline{93.37} & \textbf{3.02} & \textbf{4.73} & \textbf{25.83} & \underline{12.82} & \textbf{94.81} \\
      \midrule
      Improvement & 38.6\% & 38.2\% & 31.8\% & 41.3\% & -1.6\% & 18.0\% & 13.1\% & 22.9\% & 19.4\% & -4.1\% & 17.1\% & 16.3\% & 19.0\% & -0.3\% & 0.1\% \\
      \bottomrule
    \end{tabular}%
  }
  \vspace{1ex} \\
  \raggedright
  \footnotesize \textit{Note:} The best results are highlighted in \textbf{bold}, and the second-best results are \underline{underlined}. ''Improvement'' denotes the relative performance gain of the proposed \textit{OhmicFlow} over the best-performing baseline.
\end{table}

To assess the efficacy of the \textit{OhmicFlow} framework, we compare its overall performance against above baselines across three chronological training settings, and the results are summarized in Table \ref{tab:performance_comparison}. 

\begin{figure}
    \centering
    \includegraphics[width=0.95\linewidth,height=0.95\textheight,keepaspectratio]{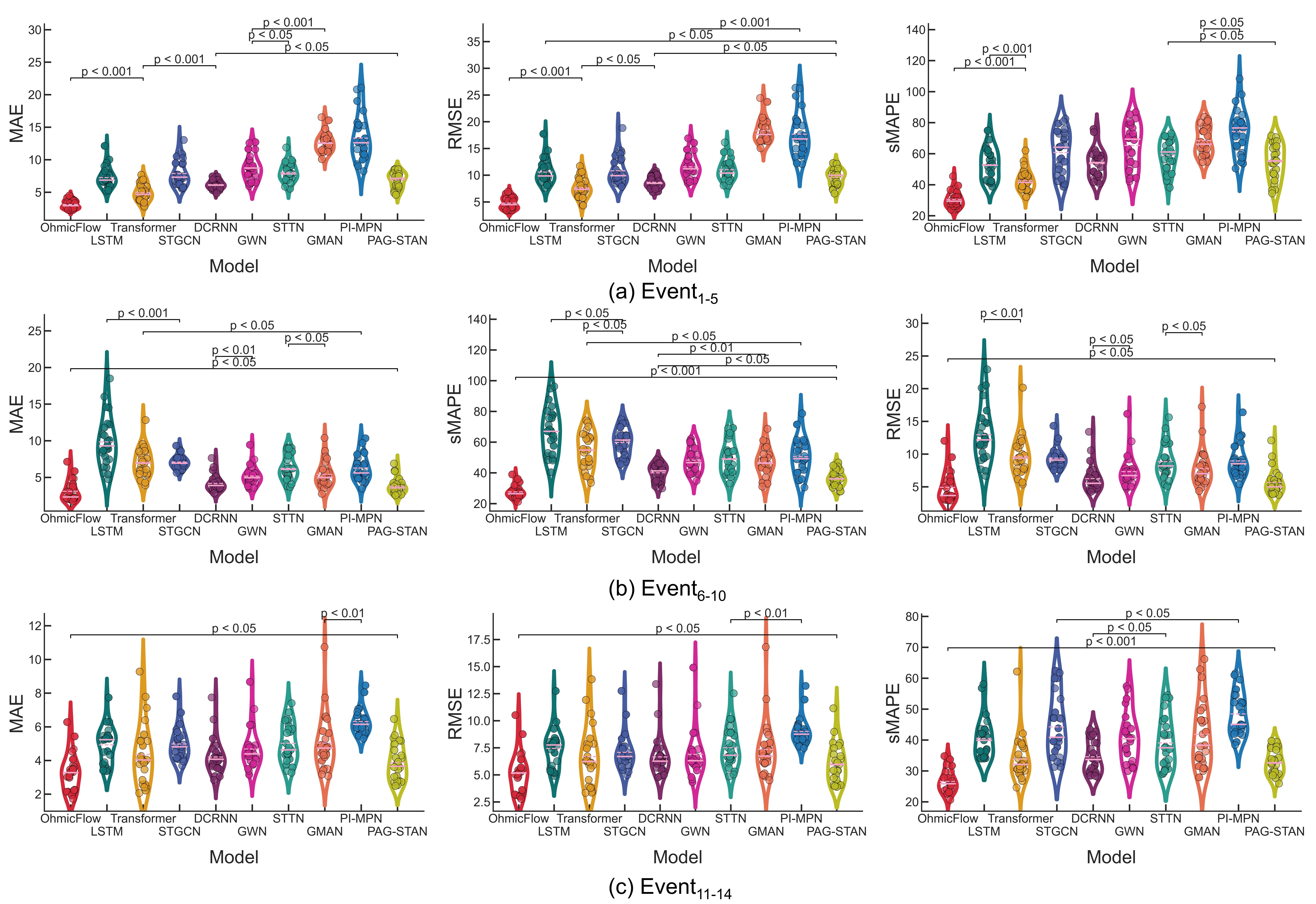}
    \vspace{-5pt}
    \captionof{figure}{Prediction errors comparison between \textit{OhmicFlow} and baselines.}
    \label{fig:violin_op}
\end{figure}

\begin{figure}
    \centering
    \includegraphics[width=0.95\linewidth,height=0.95\textheight,keepaspectratio]{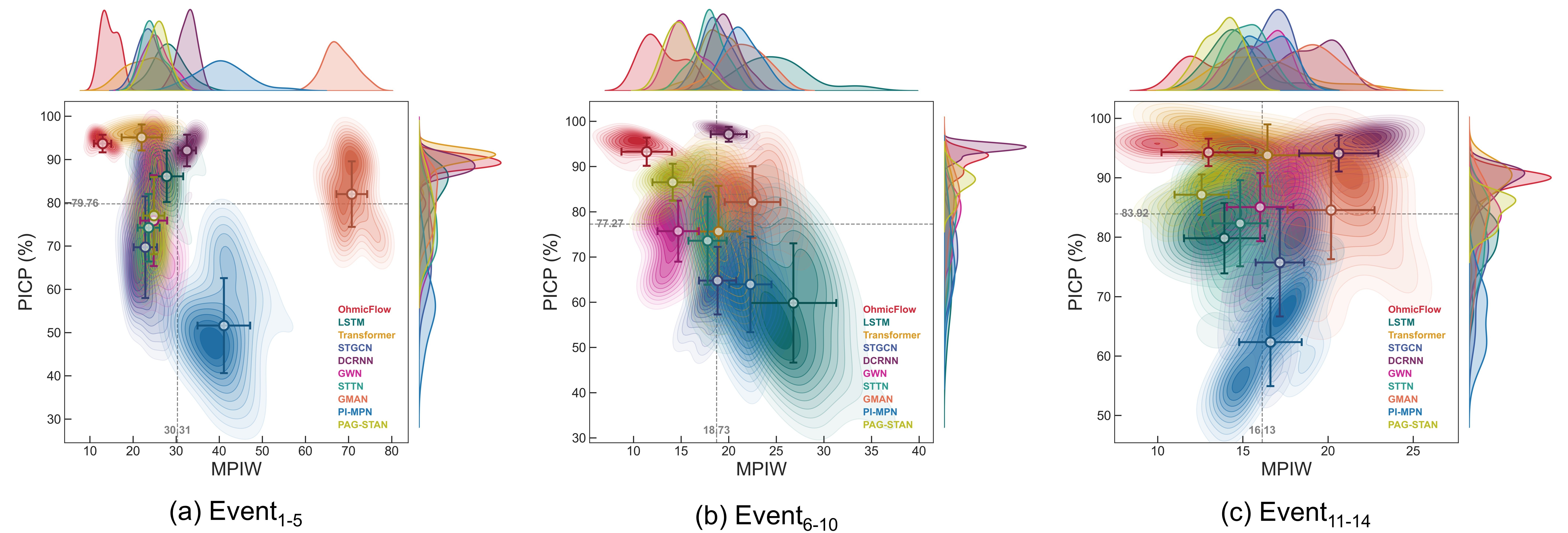}
    \vspace{-5pt}
    \captionof{figure}{Uncertainty quantification comparison between \textit{OhmicFlow} and baselines.}
    \label{fig:distribution_op}
\end{figure}

From an accuracy perspective, \textit{OhmicFlow} consistently outperforms all baselines under all three training sets. The gains are substantial when trained on the earliest events (MAE gain 38.6\%, RMSE gain 38.2\%, and sMAPE gain 31.8\%), and remain stable when trained on later 2 event sets (MAE gain 18.0\% and 17.1\%, RMSE gain 13.1\% and 16.3\%, sMAPE gain 22.9\% and 19.0\%). Fig.~\ref{fig:violin_op} complements Table~\ref{tab:performance_comparison} by showing the performance distribution, where each point corresponds to a model's accuracy on one EWE day. \textit{OhmicFlow} achieves not only the lowest average errors but also tighter dispersion, indicating more stable accuracy across heterogeneous disruptions. Its improvement over the second-best baseline is also statistically significant. From an uncertainty perspective, \textit{OhmicFlow} produces narrow intervals (MPIW: 13.40, 12.12, and 12.82) while maintaining the highest or near-highest coverage (PICP: 93.62\%, 93.37\%, and 94.81\%). Although some baselines occasionally obtain slightly higher PICP or smaller MPIW in one setting, none achieves both strong coverage and sharp intervals. This trade-off is visualized in Fig.~\ref{fig:distribution_op}, where the PICP--MPIW plane summarizes each model's uncertainty frontier: better models lie closer to the upper-left corner. \textit{OhmicFlow} occupies the most favorable upper-left region across all three training sets, and its compact spread further indicates more stable uncertainty calibration under different disruptions.

Baseline behavior also varies across training settings. Under Event$_{1-5}$, the network topology is relatively simple and spatial coupling across OD pairs remains limited. In this setting, pure temporal models are still competitive because they concentrate model capacity on dense temporal signals. Accordingly, Transformer is the strongest baseline when trained on the earliest event set. As the network expands under Event$_{6-10}$ and Event$_{11-14}$, the data size increases and most models improve, while richer connectivity makes spatial effects more important. When trained on these later events, spatiotemporal models gain more than pure temporal models. Among them, convolution-based models improve more than global attention-based models, and DCRNN achieves the best or near-best MAE and PICP among non-physics baselines. This may be because convolution-based models impose a stronger locality bias, which helps them capture stable neighborhood-level propagation and remain robust under non-stationary disruptions. In contrast, although attention mechanisms are effective for sequence modeling, their flexible global dependency modeling can make them more sensitive to disruption-induced distribution shifts and lead to higher variance when event data are limited. Against this backdrop, \textit{OhmicFlow} outperforms all baselines across training settings by combining LSTM-based temporal encoding, future-aware attention, and semantic-level GAT to capture stable temporal dynamics and focused spatial interactions.

For physics-informed baselines, PI-MPN performs relatively poorly in most settings. Besides general distribution shifts, its weak performance is likely related to the lack of POI information in our data, which reduces the directional and structural constraints required by its diffusion design. Without this component, the remaining PDE-based prior becomes less informative for disrupted transit flow dynamics. In contrast, PAG-STAN is often the second-best or best baseline, especially under Event$_{6-10}$ and Event$_{11-14}$. This suggests that conservation-based physical constraints provide useful inductive bias for disrupted flow forecasting, even when data-driven correlations become unstable. Nevertheless, \textit{OhmicFlow} achieves superior performance because its dedicated Ohmic constraint explicitly couples demand, impedance, and flow, thereby reducing implausible extrapolation and improving generalization under disruptions.

To further illustrate \textit{OhmicFlow}'s performance, Fig.~\ref{fig:spatial_op} compares two representative disrupted OD flow snapshots: 18:00--19:00 during Typhoon NIDA on 2 August 2016 and 16:00--17:00 during Typhoon SAOLA on 1 September 2023. In both cases, the OD flow patterns predicted by \textit{OhmicFlow} and PAG-STAN trained on Event$_{6-10}$ are compared with the actual disrupted flow and expected normal flow, with edge bundling used to highlight major corridor intensities. In panel (a), the metro system experienced prolonged limited operation under Typhoon NIDA, and supply contraction substantially reshaped the passenger flow. PAG-STAN's predictions remain close to the expected normal flow but deviate from the actual flow, whereas \textit{OhmicFlow} better matches the observed disruption pattern by capturing the impedance increase and reducing systematic overestimation. In panel (b), both the normal flow and PAG-STAN underestimate the actual disrupted flow, though the proposed framework again provides a closer match. At that time, Shenzhen was approaching the landfall of Typhoon SAOLA, and advance warnings had been issued for service suspension at 19:00 and storm arrival around 20:00. As a result, many passengers rescheduled their travel plans to avoid the typhoon, thereby shifting the demand peak earlier. \textit{OhmicFlow} better captures this redistribution by incorporating forthcoming severe-weather signals through future-aware attention. The results suggest that \textit{OhmicFlow} can better track disruption-induced OD flow redistribution by jointly modeling impedance changes and available future signals.

\begin{figure}
    \centering
    \includegraphics[width=0.9\linewidth,height=0.9\textheight,keepaspectratio]{figs/Spatial_OP.jpg}
    \vspace{-5pt}
    \captionof{figure}{Comparison of spatial patterns among the actual disrupted flow, expected normal flow, predicted flow by \textit{OhmicFlow} and predicted flow by a baseline model.}
    \label{fig:spatial_op}
\end{figure}

\subsection{Ablation analysis}

To assess how much \textit{OhmicFlow} benefits from its circuit-inspired physical constraints, we conduct an ablation study by removing each constraint and comparing the resulting variants. We focus on two key constraints: Ohmic consistency and KCL-inspired inflow conservation supervision. Therefore, we test three variants:

\begin{itemize}[leftmargin=\parindent,labelsep=0.5em,noitemsep]
    \item \textbf{\textit{OhmicFlow w/o Ohm}} removes the voltmeter branch and the PTC module, and predicts only flow $I$ with the ammeter branch. This variant does not perform joint inference of demand $U$ and impedance $R$, but still keeps the KCL-inspired origin-centric inflow supervision. Its training objective is $\mathcal{L}_{\mathrm{od}}+\lambda_{\mathrm{in}}\mathcal{L}_{\mathrm{in}}$.
    \item \textbf{\textit{OhmicFlow w/o KCL}} keeps the Ohmic constraint but removes the KCL-inspired inflow conservation supervision, i.e., no supervision on aggregated station inflow. Its objective is $\mathcal{L}_{\mathrm{od}}+\mathcal{L}_{\mathrm{ohm}}+\mathcal{L}_{\mathrm{soft}}$, without $\mathcal{L}_{\mathrm{in}}$.
    \item \textbf{\textit{OhmicFlow-Vanilla}} removes both the Ohmic and KCL constraints and trains only with $\mathcal{L}_{\mathrm{od}}$.
\end{itemize}

\begin{figure}
    \centering
    \includegraphics[width=0.95\linewidth,height=0.95\textheight,keepaspectratio]{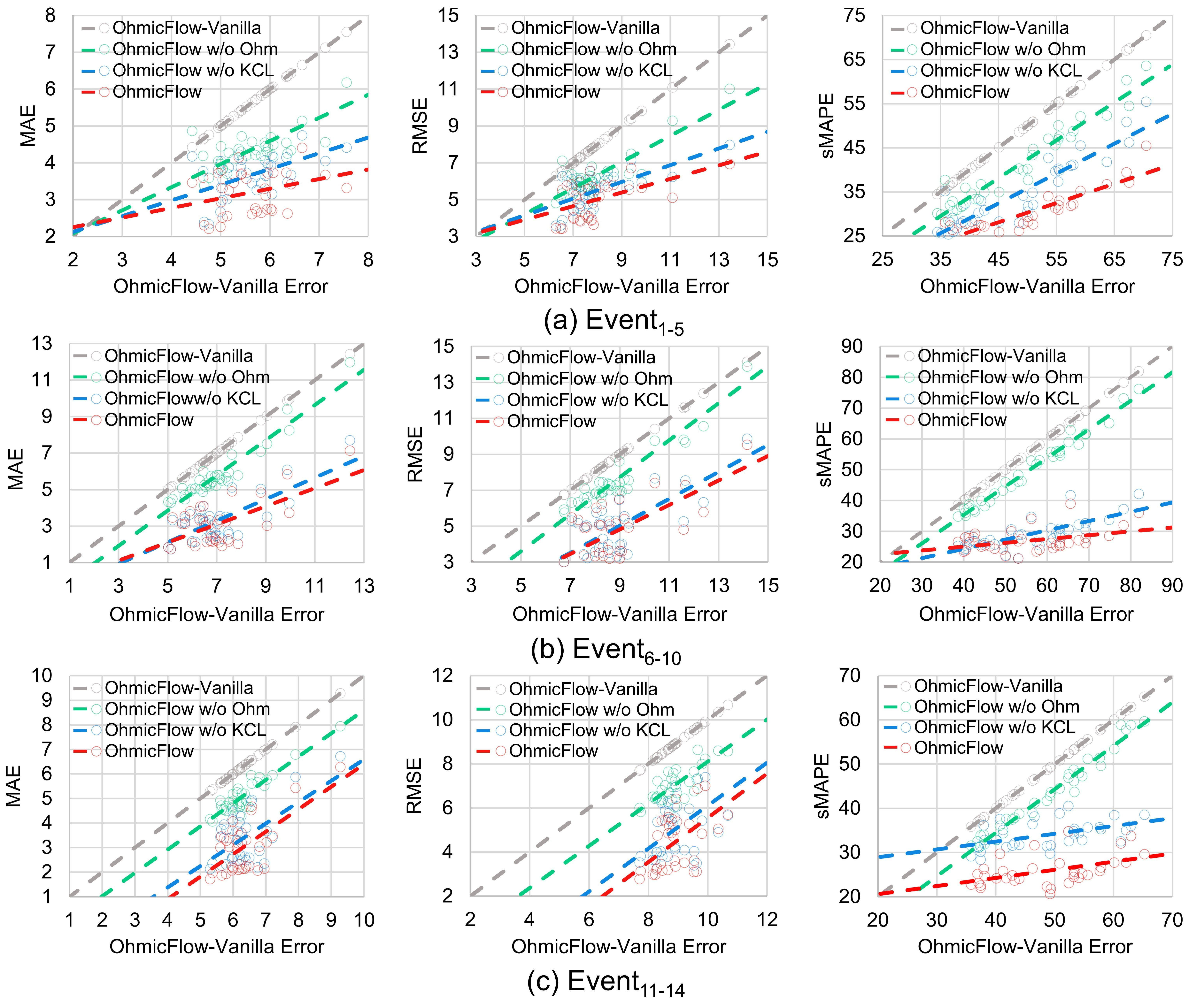}
    \vspace{-5pt}
    \captionof{figure}{Prediction errors of ablation variants mapped against the OhmicFlow-Vanilla (y=x).}
    \label{fig:curve_fit_ab}
\end{figure}

Fig.~\ref{fig:curve_fit_ab} evaluates model accuracy using paired error plots against \textit{OhmicFlow-Vanilla}. Each point compares the prediction error of the vanilla model on one EWE day (x-axis) with the error of another variant on the same day (y-axis), with regression lines fitted to show overall trends. Lines below $y=x$ indicate improvements over \textit{OhmicFlow-Vanilla}. Since larger errors indicate lower predictability under the vanilla model, the x-axis also serves as a proxy for event-level prediction difficulty. Across training settings and metrics, removing either the Ohmic or KCL constraint shifts the trend lines upward relative to \textit{OhmicFlow}, indicating accuracy degradation; equivalently, adding either constraint to \textit{OhmicFlow-Vanilla} shifts the lines downward, confirming consistent gains. The Ohmic constraint contributes more significantly, as the trend line of \textit{OhmicFlow w/o Ohm} lies above that of \textit{OhmicFlow w/o KCL} in almost all cases. Notably, \textit{OhmicFlow} and \textit{OhmicFlow w/o KCL} have slopes clearly below 1, suggesting that the Ohmic constraint brings larger absolute gains when disrupted events are harder to predict. The figure also shows complementarity: the KCL constraint improves prediction, while adding the Ohmic constraint provides further gains.

\begin{figure}
    \centering
    \includegraphics[width=0.95\linewidth,height=0.95\textheight,keepaspectratio]{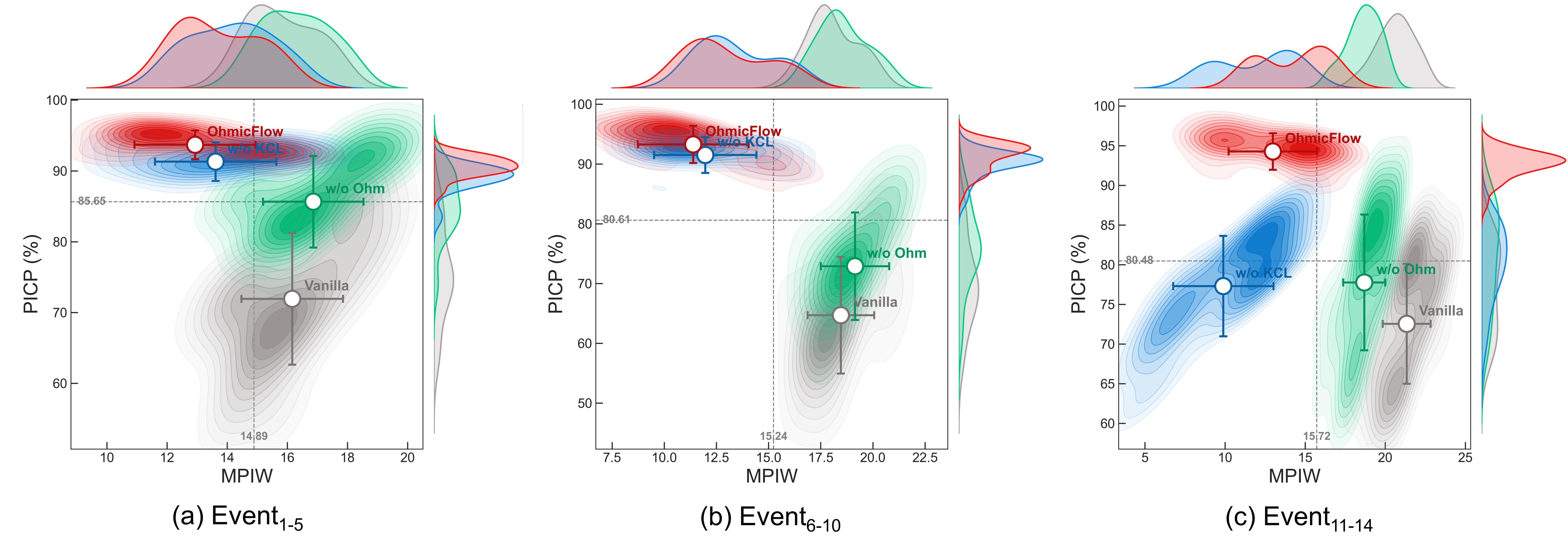}
    \vspace{-5pt}
    \captionof{figure}{Uncertainty quantification comparison between ablation variants.}
    \label{fig:distribution_ab}
\end{figure}

Fig.~\ref{fig:distribution_ab} analyzes prediction uncertainty using the PICP--MPIW plane. The horizontal and vertical dashed lines mark the average PICP and MPIW across all variants and all EWE days, dividing the space into four quadrants. Removing the physical constraints generally worsens uncertainty calibration. The effect is much stronger when the Ohmic constraint is removed: \textit{OhmicFlow w/o Ohm} shifts from the favorable upper-left region to the unfavorable lower-right region, whereas \textit{OhmicFlow w/o KCL} remains in the upper-left or lower-left region. This suggests that the Ohmic constraint is the main driver of uncertainty control, while the KCL constraint provides an additional but secondary stabilizing effect.

\subsection{Extensibility of the \textit{OhmicFlow} framework}

As mentioned earlier and illustrated in Fig.~\ref{fig:most}, \textit{OhmicFlow} is designed as a modular framework in which the spatiotemporal sensing backbone can be replaced while the circuit-inspired physical components are kept unchanged. In our default setting, TFT-GAT serves as this backbone and, when coupled with the unified embeddings, the voltmeter bypass, the PTC impedance module, and the Ohmic constraint, delivers superior overall prediction performance in terms of both accuracy and uncertainty compared to the baselines. However, as better data-driven backbones emerge in the future, it is essential to ensure the extensibility of the proposed framework. To this end, we replace TFT-GAT with each baseline introduced in Sec.~\ref{sec:baselines}, while keeping the unified embeddings, voltmeter bypass and PTC impedance module unchanged, and evaluate the efficacy of model transplantation.

\begin{figure}
    \centering
    \includegraphics[width=0.95\linewidth,height=0.95\textheight,keepaspectratio]{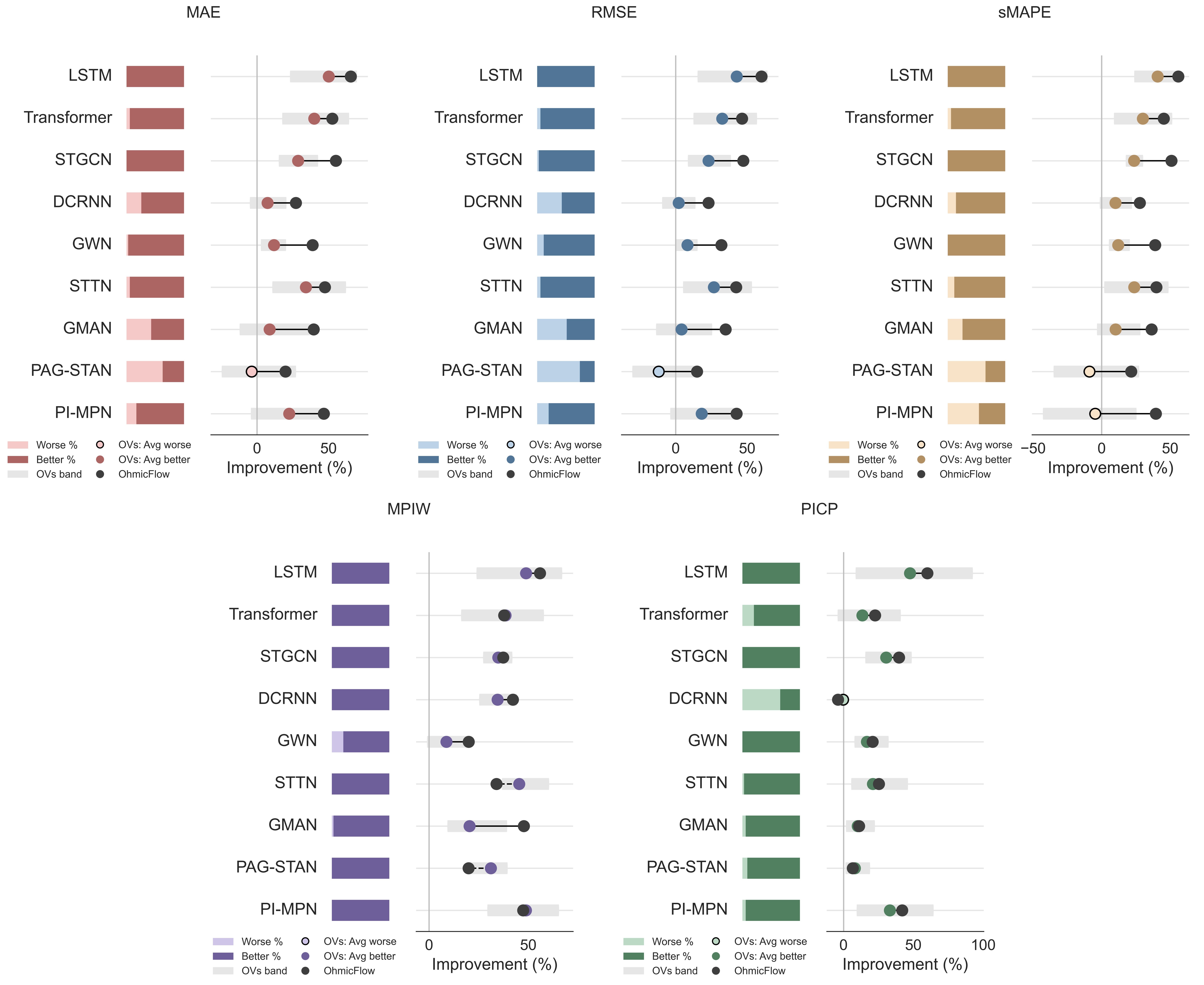}
    \vspace{-5pt}
    \captionof{figure}{Performance improvements of Ohmic variants over standard baselines.}
    \label{fig:extensibility}
\end{figure}

Fig.~\ref{fig:extensibility} summarizes the relevant findings. For visual clarity, we exclusively present the results trained on Events$_{6-10}$. Herein, we refer to the transplanted models as \textit{Ohmic Variants} (OVs). The left stacked bars delineate the proportion of test EWE days where each OV either outperforms or underperforms its original counterpart. Concurrently, the right bubble chart quantifies the average relative improvements over the original baselines. Specifically, dark bubbles denote positive gains, whereas lightly shaded bubbles with black borders indicate no improvement. Furthermore, solid black bubbles represent the relative improvements of our native \textit{OhmicFlow} (utilizing a TFT-GAT backbone) against each baseline, serving as a reference to judge whether OVs can surpass the default architecture.

Overall, model transplantation under the \textit{OhmicFlow} framework improves both accuracy and uncertainty for most backbones. The only notable exception is PAG-STAN, whose already strong baseline performance and high architectural complexity likely leave limited room for additional accuracy gains after transplantation; correspondingly, its OV shows slight accuracy degradation. Even so, PAG-STAN's OV still yields a clear reduction in predictive uncertainty. More broadly, although most OVs obtain positive average gains, they rarely exceed the native TFT-GAT-based architecture, especially in terms of accuracy. This pattern indicates that the \textit{OhmicFlow} framework is broadly transferable, and TFT-GAT is particularly well suited for the imposed Ohmic constraint.

\subsection{Robustness analysis}\label{sec:robustness}

Robustness is important because \textit{OhmicFlow} is intended for disrupted transit networks, where the model performance should remain reliable not only on average but also across heterogeneous OD types and imperfect forward-looking signals. We therefore examine robustness from two complementary perspectives: first, whether the model delivers consistent gains across OD groups with different flow characteristics; and second, how sensitive it is to errors in the forward-looking inputs used to characterize disruptions.

\begin{figure}
    \centering
    \includegraphics[width=0.95\linewidth,height=0.95\textheight,keepaspectratio]{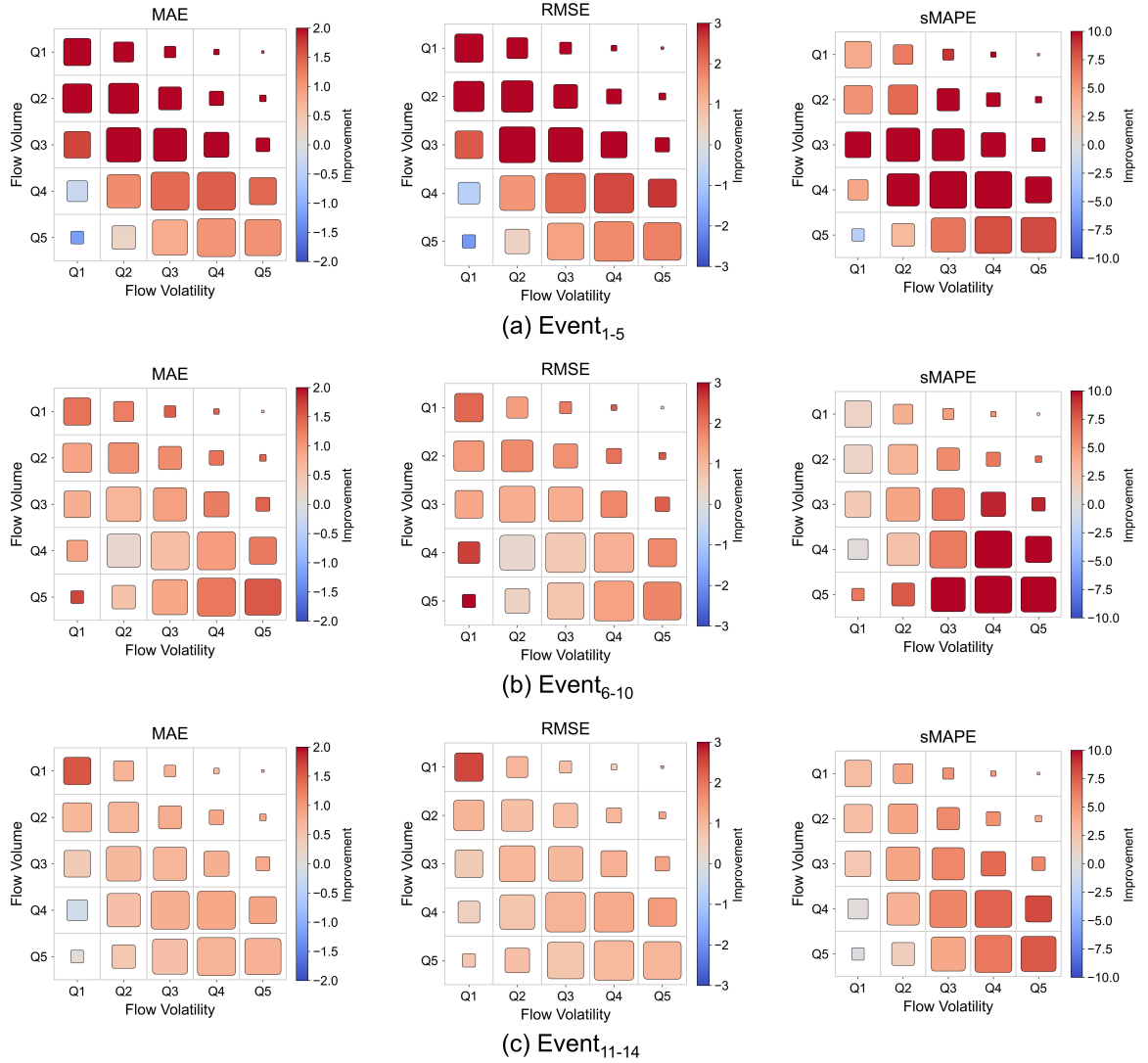}
    \vspace{-5pt}
    \captionof{figure}{Performance improvements across OD groups partitioned by flow volume and volatility.}
    \label{fig:partition}
\end{figure}

For the first part, we partition OD pairs into a $5\times5$ matrix according to flow volume and flow volatility, yielding 25 groups. Flow volume is measured by the total actual flow in the future window, i.e., $\sum_{t\in\mathcal{T}_f} I_{k,t}$, while flow volatility is defined by the absolute discrepancy between the actual disrupted flow and expected normal flow, i.e., $|\sum_{t\in\mathcal{T}_f} I_{k,t}-\sum_{t\in\mathcal{T}_f} I^b_{k,t}|$. This partition stratifies OD pairs by flow scale and disruption-induced flow deviation, allowing us to verify \textit{OhmicFlow}'s consistent accuracy across OD pairs with different operational characteristics. In Fig.~\ref{fig:partition}, the rectangle color indicates the accuracy gain of \textit{OhmicFlow} over \textit{OhmicFlow-Vanilla}, with red and blue denoting positive and negative gains, respectively, while the rectangle size reflects the number of OD pairs in each partition. Q1 represents the top partition, which captures the upper 20\% cumulative volume or volatility. Across all training settings, \textit{\textit{OhmicFlow}} achieves positive gains in nearly all partitions. The few exceptions appear mostly in low volume but highly volatile groups, where sparse observations and noisy patterns make the data less informative; this effect is most visible in Event$_{1-5}$ when training data are limited. Crucially, \textit{OhmicFlow} consistently yields positive gains not only in the most critical high-volume/high-volatility partitions, but also across stable, low-volatility groups. Together, these patterns suggest that \textit{OhmicFlow} stays robust across OD types: the Ohmic constraint couples demand, impedance, and flow, making predictions less sensitive to sparse noise and abnormal conditions.

For the second part, we examine perturbations on the forward-looking inputs explicitly modeled by \textit{OhmicFlow}. These covariates, namely $W_t$ and $\tau_{k,t}$, represent disruption-related drivers of demand fluctuations and impedance changes. In practice, however, they can be imperfect: weather forecasts can be inaccurate, operational plans may be revised, and $\tau_{k,t}$ is rarely known exactly. Instead, future travel time is often approximated by matching planned operations or future events with historical travel time patterns under similar disruption conditions and service arrangements. To mimic such uncertainty, we design three perturbation settings. For wind- and rain-related variables, we consider: (i) a mutation scheme that randomly adds or subtracts one severity level at each time step with a specified probability to simulate intensity forecast errors; and (ii) a temporal shift scheme that shifts the entire weather sequence forward or backward to simulate timing misalignment, such as earlier or later typhoon landfall. This setup injects random noise while preserving the overall weather trend. For travel time, we inject OD-specific multiplicative noise within $[-r,r]$, i.e., $\tau_{k,t}$ is scaled by $(1+r)$ while remaining no smaller than the normal travel time, so that the perturbation changes its magnitude but preserves its main trend. This setting reflects the practical situation where operators can infer an approximate efficiency range rather than the exact travel time.

\begin{figure}
    \centering
    \includegraphics[width=0.9\linewidth,height=0.9\textheight,keepaspectratio]{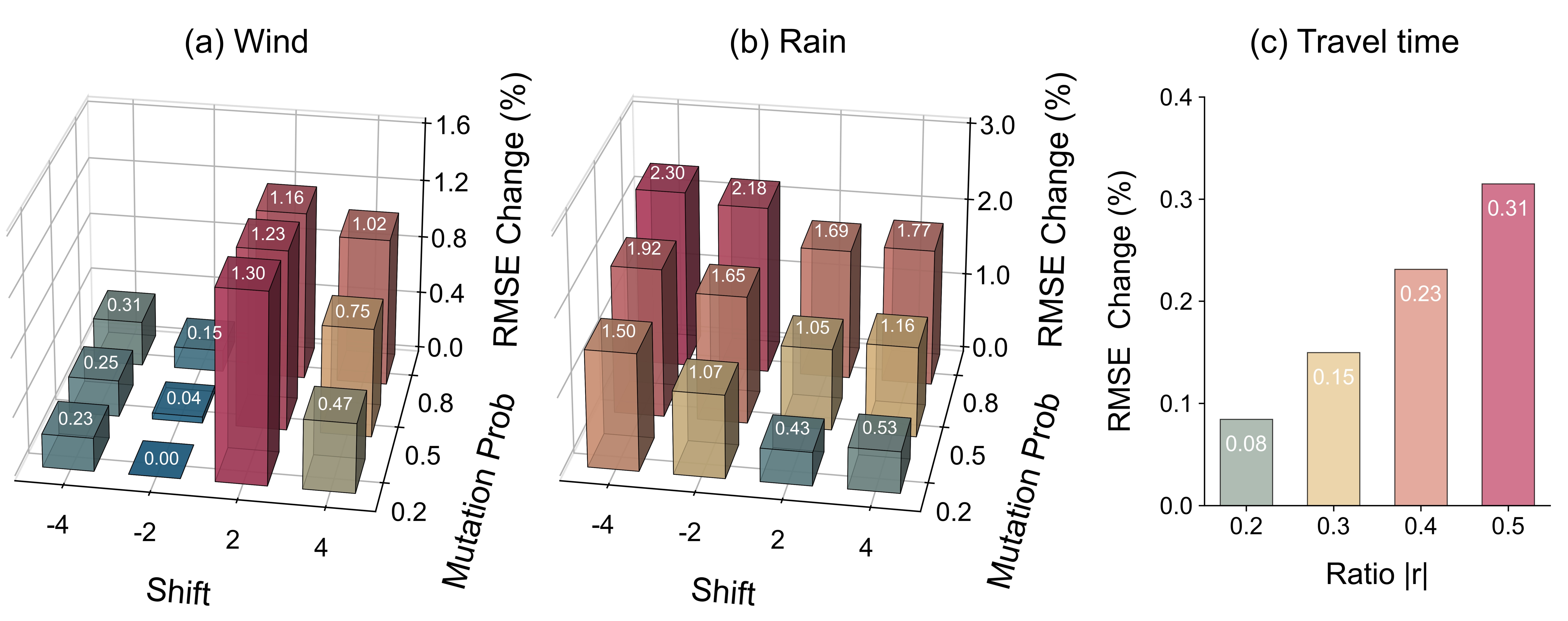}
    \vspace{-5pt}
    \captionof{figure}{Model sensitivity to different perturbations in wind, rain, and travel time.}
    \label{fig:perturb}
\end{figure}

Fig.~\ref{fig:perturb} reports the resulting performance degradation using the relative change in RMSE, which is more sensitive to outliers than MAE. \textit{OhmicFlow} here is also trained on Event$_{6-10}$. As expected, higher mutation probabilities for both wind and rain generally lead to larger degradation. Notably, \textit{OhmicFlow} is more sensitive to delayed typhoon warnings and to advanced rainstorm warnings. This asymmetry aligns with their distinct disruption mechanisms: typhoon effects are often anticipatory, with demand shifting before landfall due to service suspensions and evacuations; conversely, rainfall impacts are primarily reactive, characterized by physical blockages (e.g., waterlogging) emerging after the onset. Temporally misaligning these signals exposes a larger proportion of the future window to more misleading disruption evidence. The model's logical sensitivity to such timing errors verifies its accurate representation of real-world weather mechanisms. Despite this sensitivity, the maximum degradation remains modest at 2.30\%, which is acceptable for flow forecasting under disruptions. For travel time, larger random noise inevitably causes larger degradation, but the maximum change is still below 0.31\%. This robustness may stem from \textit{OhmicFlow}'s ability to effectively absorb localized point-wise perturbations. By leveraging semantic-level GAT operations and capturing spatiotemporal spillovers of impedance, the framework fundamentally prioritizes overarching trends over exact point values. Overall, these results show that \textit{OhmicFlow} is robust to both OD heterogeneity and moderate perturbations in forward-looking inputs, because its physical constraints stabilize the mapping from disruption signals to flow predictions.

\subsection{Explainability}

\begin{figure}
    \centering
    \includegraphics[width=0.9\linewidth,height=0.9\textheight,keepaspectratio]{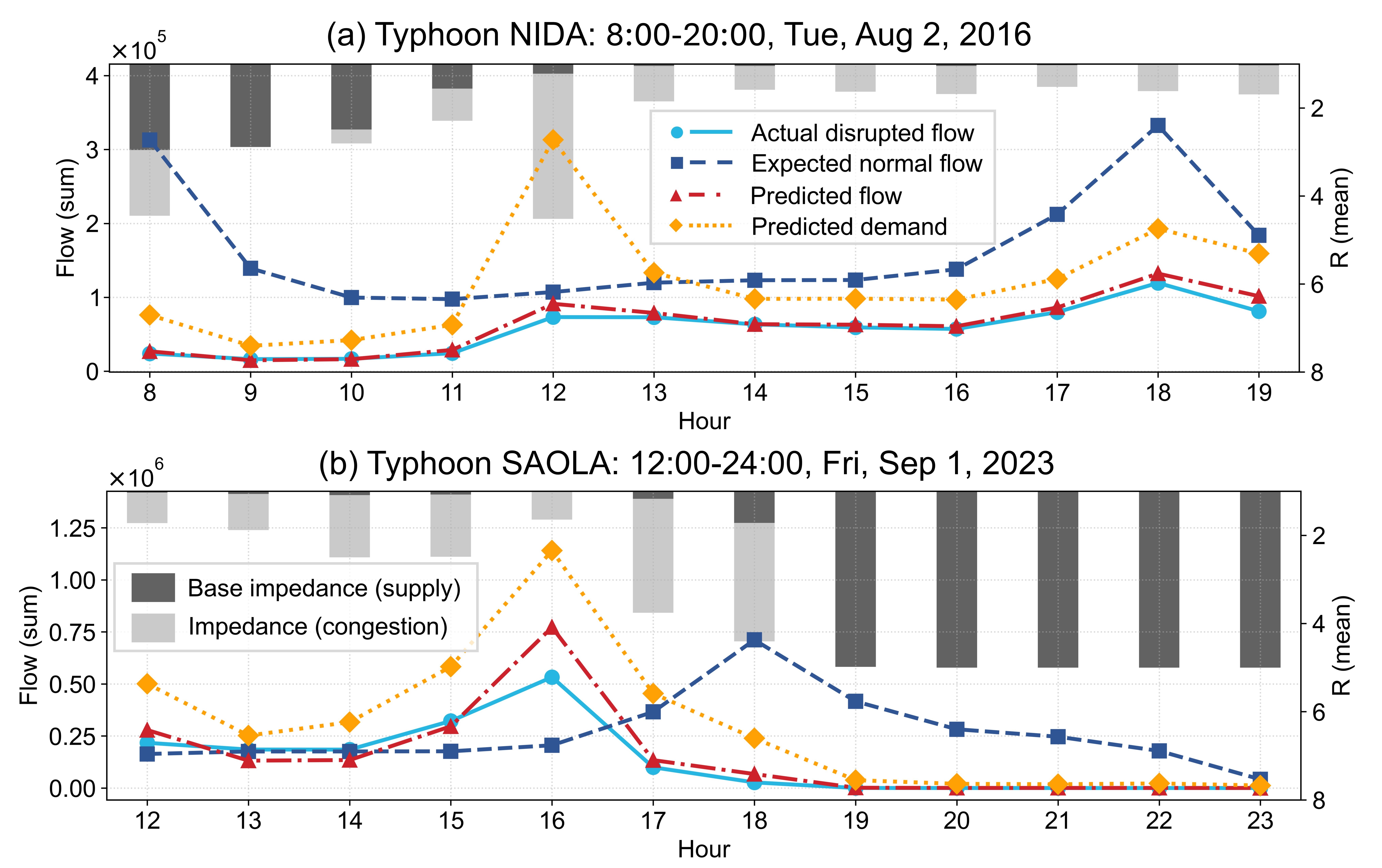}
    \vspace{-5pt}
    \captionof{figure}{Interpretable Dynamics of Flow Prediction under Varying Demand and Impedance.}
    \label{fig:explain_line}
\end{figure}

\begin{figure}
    \centering
    \includegraphics[width=0.9\linewidth,height=0.9\textheight,keepaspectratio]{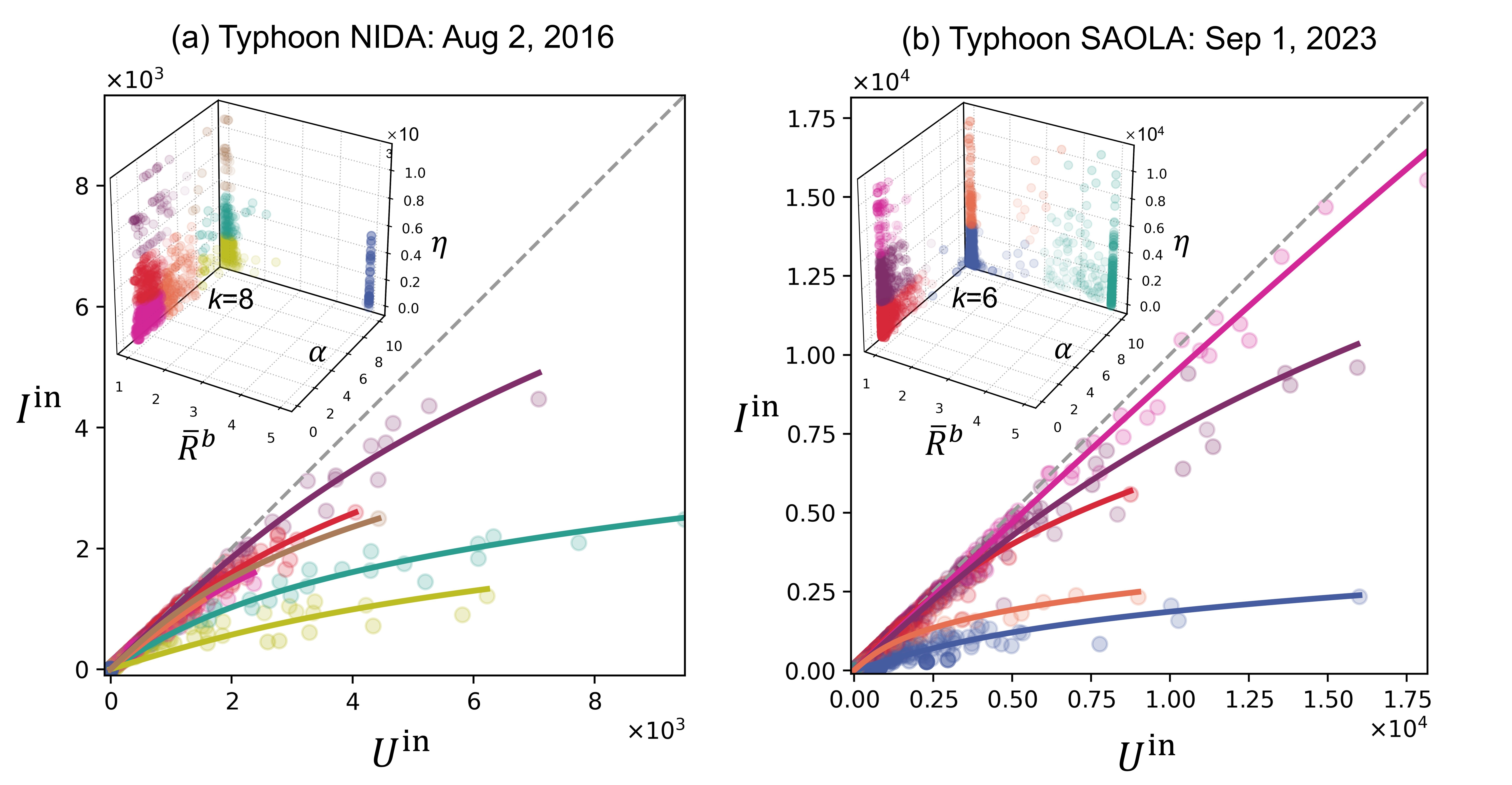}
    \vspace{-5pt}
    \captionof{figure}{Heterogeneous Non-linear Flow-Demand Relationships under Impedance Constraints.}
    \label{fig:explain_iu}
\end{figure}

Beyond performance gains, \textit{OhmicFlow} also provides built-in mechanistic explainability. Rather than predicting passenger flow as an isolated target, it reveals the covariation among flow, latent demand, and travel impedance. The intermediate physical parameters produced by the PTC module further expose heterogeneous impedance patterns across OD pairs, offering an analytical lens for evaluating the resilience of transit systems under disruption.

We first examine how \textit{OhmicFlow} disentangles whether flow changes are mainly driven by demand or impedance. Fig.~\ref{fig:explain_line} presents two 12-hour cases predicted by \textit{OhmicFlow}, trained on Event$_{6-10}$: 08:00--20:00 during Typhoon NIDA on 2 August 2016 and 12:00--24:00 during Typhoon SAOLA on 1 September 2023. For each case, we compare system-level trajectories of predicted flow, latent demand, and average impedance, together with the actual disrupted flow and expected normal flow as a reference). In panel (a), the network remains under limited operation before 12:00 due to lingering impacts of NIDA. Insufficient supply dominates impedance, while demand also stays below the normal flow, jointly reducing flow. Around 12:00, full-line service resumes and previously suppressed demand is released, but the surge in latent demand does not fully convert to passenger flow as congestion becomes the new dominant impedance. Later, as demand weakens, the flow remains below the reference even after the impedance declines. Essentially, lower than usual passenger flow within a 12-hour period is the result of various mechanisms. Panel (b) shows another pathway: around 16:00, SAOLA landfall warnings and service suspension notices shift the evening demand peak earlier, causing passenger flow to exceed the reference and then increasing congestion-induced impedance over the next 1-2 hours. After 18:00, sharp demand contraction and worsening supply jointly push the flow close to zero. These results show that \textit{OhmicFlow} can jointly infer the demand and impedance dynamics behind observed flow, helping operators distinguish between the needs for demand management and capacity allocation under disruptions.

Next, we show how intermediate physical outputs from the PTC module characterize the resilience structure of a disrupted transit network. Given an \textit{OhmicFlow} trained on observed events, such as Event$_{6-10}$, the PTC parameters $R^b_{k,t}$ and $\alpha_{i,t}$ quantify the OD impedance characteristics when new network descriptors $(f_{i,p},f_{k,p},\mathcal{N}_{p,k},\eta_{i,p})$ and anticipated disruption signals $(W_t,\tau_{k,t})$ are provided. For consistent parameter granularity, we aggregate OD-level base impedance by the origin station as $\bar{R}^{b}_{i,t}={1}/{|V_p|} \times \sum_{j \in V_p}R^{b}_{k=(i,j),t}$, and then approximately map the Ohmic relation to the station level: $U_{i,t}^{\mathrm{in}}/I_{i,t}^{\mathrm{in}}\approx \bar{R}^{b}_{i,t}[1+\alpha_{i,t}(I^{\mathrm{in}}_{i,t}/\eta_{i,p})^2]$. Fig.~\ref{fig:explain_iu} illustrates the same two cases as Fig.~\ref{fig:explain_line} in the $I^{\mathrm{in}}$--$U^{\mathrm{in}}$ space, with each point representing an origin station at a specific hour. We then cluster these points by $\bar{R}^{b}_{i,t},\alpha_{i,t}$ and $\eta_{i,p}$ using k-means, and fit cluster-specific trends to obtain impedance characteristic curves.

The resulting curves can be interpreted as a resilience structure of the disrupted network, showing which OD impedance patterns dominate under disruption. In both panels, flow increases sublinearly with demand and asymptotically reaches a congestion-limited saturation bound, reflecting diminishing marginal flow conversion under heavy demand pressure. The saturation rate varies among component types: a higher $\bar{R}^{b}_{i,t}$ suppresses the initial flow conversion, while a larger $\alpha_{i,t}$ and smaller $\eta_{i,p}$ lead to earlier and stronger saturation, collectively reflecting weaker resilience. Meanwhile, network expansion improves transport capacity and redundancy, so some components maintain higher flow conversion even under high demand. This is visible in panel (b), where some of the OD groups (pink curve) saturates more slowly than in panel (a), despite SAOLA being more severe than NIDA. However, expansion also amplifies heterogeneity: some components in panel (b) representing more vulnerable OD groups (blue curve) saturate earlier. Therefore, \textit{OhmicFlow} not only explains predicted flow trajectories but also uncovers the resilience status across OD pairs, thereby supporting targeted interventions under EWEs.

\section{Conclusion}\label{sec:conclusion}
This study presents a novel perspective to conceptualize the transit system as an electrical circuit and, inspired by Ohm's law, proposes the \textit{OhmicFlow} framework for forecasting passenger flow under disruption. Rather than treating flow prediction as a single-target regression task, \textit{OhmicFlow} models the coupled covariation among passenger flow, latent demand, and travel impedance. It consists of three coordinated modules: an ammeter-equivalent learner that predicts disrupted passenger flow using future-aware TFT and semantic-level GAT; a voltmeter-inspired counterfactual bypass that infers latent demand through a shared-parameter ammeter replica under impedance-controlled inputs; and a PTC-informed impedance module that distinguishes between supply contraction and congestion effects to estimate dynamic travel impedance. Experiments on Shenzhen Metro using 10 years of data and 17 EWEs show consistent improvements in point accuracy and uncertainty quality. Ablation results further confirm that the Ohmic constraint is the primary contributor, while KCL-inspired inflow conservation provides complementary stabilization.

Beyond predictive performance, \textit{OhmicFlow} offers direct operational value for emergency management. By learning transferable disruption knowledge from historical weather events, a trained model can be deployed before a forthcoming unseen disruption, using forward-looking signals to generate point and interval forecasts for emergency response planning. Since \textit{OhmicFlow} explicitly considers anticipated weather states and anticipated disrupted travel time, while remaining relatively insensitive to their input errors, it is well suited for uncertain pre-event deployment. The required inputs are practical and widely available, including passenger flow records (e.g., AFC records), transit network topology, and public weather information, making the framework transferable across cities and extensible to other foreseeable disruptions, such as large-scale public events \citep{Liang2024Exploring}. \textit{OhmicFlow} also supports what-if stress testing by adjusting weather forecasts and planned travel time, allowing transit agencies to simulate flow evolution under alternative disruption and operation scenarios; thus, it serves not only as a predictor but also as a decision-support testbed. The jointly inferred demand and impedance explain why flow changes, helping operators decide whether to prioritize demand management or supply adjustment. Meanwhile, physical outputs such as station congestion sensitivity and base impedance reveal the system resilience structure under current network states and anticipated disruptions, enabling targeted interventions for OD pairs with weak demand-to-flow conversion. With a modular architecture, \textit{OhmicFlow} can also be attached to different spatiotemporal backbones, ensuring the adaptability to transit agencies' existing systems.

This study can be further improved and extended in several directions. First, due to data availability, we currently use travel time, derived from passenger flow records, as a proxy for operational disturbance and as an input for modulating the physical quantities in the PTC module. In practice, transit operators also maintain train/bus running plans and dynamic timetables, which provide more direct operational variables to capture supply contraction and potentially improve impedance modeling. Second, because Ohmic and KCL constraints are introduced through a multi-objective loss, the weights of data-driven and physics-informed terms are currently set empirically based on preliminary studies. Although this generic setting already yields strong performance, recent PINN studies have proposed adaptive weighting schemes to seek Pareto-optimal trade-offs between data and physical objectives \citep{Lei2025Reconstructing}; such methods could be incorporated to further improve \textit{OhmicFlow}. Third, \textit{OhmicFlow} currently supports forecasting and scenario probing rather than closed-loop control. A key next step is to integrate it with a control optimization module, so that predictions can directly inform feasible operational policies, such as dynamic timetabling, crowd management, and resource allocation under explicit resilience objectives. In this setting, future-aware inputs to \textit{OhmicFlow}, such as $\tau_{k,t}$, can be updated based on the feedback from selected control actions, allowing the framework to iteratively guide the network toward a more stable state.

\section*{Acknowledgment}
This research is supported by the Natural Science Fund of Guangdong Province (2026A1515011818) and Seed Fund for Basic Research at The University of Hong Kong (109000301).

\printcredits

\bibliographystyle{cas-model2-names}

\FloatBarrier
\bibliography{ref}



\end{document}